\documentclass[aps,prc,preprint,tightenlines,floatfix,groupedaddress,
nofootinbib,showpacs,preprintnumbers,amsmath,amssymb,superscriptaddress]{revtex4}
\usepackage{graphicx}
\usepackage{dcolumn}
\usepackage{mathrsfs}
\usepackage{bm}

\usepackage{graphicx}
\usepackage{dcolumn}
\usepackage{bm}
\usepackage[usenames]{color}

\begin{document}

\title{Sensitivity of the Moment of Inertia of Neutron
         Stars to the Equation of State of Neutron-Rich 
         Matter}

\author{F.~J.~Fattoyev}
\email{ff07@fsu.edu}
\author{J.~Piekarewicz}
\email{jpiekarewicz@fsu.edu}
\affiliation{Department of Physics, Florida State University, 
                  Tallahassee, Florida 32306, USA}

\date{\today}

\begin{abstract}
The sensitivity of the stellar moment of inertia to the neutron-star
matter equation of state is examined using accurately-calibrated
relativistic mean-field models. We probe this sensitivity by tuning
both the density dependence of the symmetry energy and the high
density component of the equation of state, properties that are at
present poorly constrained by existing laboratory data.  Particularly
attractive is the study of the fraction of the moment of inertia
contained in the solid crust. Analytic treatments of the crustal
moment of inertia reveal a high sensitivity to the transition pressure
at the core-crust interface. This may suggest the existence of a
strong correlation between the density dependence of the symmetry
energy and the crustal moment of inertia. However, no correlation was
found. We conclude that constraining the density dependence of the
symmetry energy---through, for example, the measurement of the 
neutron skin thickness in ${}^{208}$Pb---will place no significant
bound on either the transition pressure or the crustal moment of
inertia.
\end{abstract}
\maketitle

\section{Introduction}

Since their discovery more than 40 years ago, neutron stars have been
a fruitful testing ground for both general relativity and theories of
dense matter. Being extremely compact (with typical compactness
parameters of $R_{\rm s}/R\!\equiv\!2GM/c^2R\!\sim\!0.4$) and
extremely dense (with densities exceeding several times nuclear matter
saturation density) neutron stars are unique laboratories for the
study of exotic phenomena that lie well outside the realm of
terrestrial laboratories. Moreover, neutron stars provide a natural
meeting place for nuclear physics and astrophysics. And with the
advent and commissioning of powerful facilities dedicated to the
production of rare isotopes, new telescopes operating at a variety of
wavelengths, and more sensitive gravitational wave detectors, the
partnership between nuclear physics and astrophysics will only
continue to grow.

In the present contribution we study the sensitivity of the stellar
moment of inertia to the underlying equation of state. Although the
formalism is general---at least within the {\sl slow-rotation
approximation}---we focus on the recently discovered binary pulsar PSR
J0737-3039~\cite{Burgay:2003jj,Lyne:2004cj}.  Located ten times closer
than the celebrated Hulse-Taylor binary system~\cite{Hulse:1974eb} and
with the shortest orbital period of its kind (almost three times
smaller than the Hulse-Taylor binary), PSR J0737-3039 is the first
ever discovered double pulsar.  This discovery has resulted in some of
the most precise tests of Einstein's general theory of relativity to
date. Moreover, it has also enabled the accurate determination of
several pulsar properties, such as the orbital period of the binary, 
both pulsar masses, and both spin periods. However, their
individual radii, moments of inertia, gravitational redshifts, or any
other property that would allow a simultaneous mass-radius
determination---and therefore place important constraints on the
equation of state---are still unavailable.  Yet the prospects for
measuring the moment of inertia of the fastest spinning pulsar in the
binary (PSR J0737-3039A) have never been better. Doing so with 
even a 10\% accuracy may provide stringent constraints on the
nuclear equations of
state~\cite{Morrison:2004df,Lattimer:2004nj,Bejger:2005jy,Lavagetto:2006ew}.
Note, however, that the proposed 10\% accuracy has been recently put
into question~\cite{Iorio:2007yz}. Yet we trust that the various
observational challenges will be met successfully in the near
future. With such a view in mind, we focus on a particular feature of
the equation of state (EOS) that has a strong impact on the moment 
of inertia: {\sl the nuclear symmetry energy}.

The nuclear symmetry energy represents the increase in the energy of
the system as protons are converted into neutrons (or neutrons into
protons).  Whereas ground-state masses constrain the symmetry energy
near saturation density, they leave its density dependence
(slope, curvature, {\sl etc.}) largely undetermined. This is important
as the slope of the symmetry energy at saturation density determines
the pressure of pure neutron matter (PNM).  And it is precisely
the pressure of PNM that provides the necessary stellar
support against gravitational collapse. Thus, the larger the pressure
of PNM the larger the radius of the neutron star. Note that it is also
the pressure of PNM that accounts for the size of the neutron skin
thickness in medium to heavy nuclei. As a result, a strong correlation
was uncover between the neutron skin thickness of ${}^{208}$Pb and the
neutron radius of a neutron star~\cite{Horowitz:2001ya}. Given that
the moment of inertia scales as the square of the stellar radius,
the aim of this work is to study the expected correlation between the
neutron skin thickness of ${}^{208}$Pb and the stellar moment of
inertia. We should mention that at the time of this writing the
experimental effort to measure the neutron radius of ${}^{208}$Pb at
the Jefferson Laboratory was well on its way. The Parity Radius
Experiment (``{\sl PREx}") at the Jefferson Laboratory aims to measure
the neutron radius of $^{208}$Pb accurately and model independently
via parity-violating electron
scattering~\cite{Horowitz:1999fk,Michaels:2005}. PREx should provide a
unique experimental constraint on the neutron skin thickness of a
heavy nucleus and correspondingly on the pressure of pure neutron
matter.  We should also mention that considerable progress has
been achieved in the theoretical front.  By building on the universal
behavior of dilute Fermi gases with an infinite scattering length,
significant theoretical progress has been made in constraining the
equation of state of pure neutron matter~\cite{Baker:1999dg,
Heiselberg:2000bm,Carlson:2003,Schwenk:2005ka,Nishida:2006br,
Hebeler:2009iv, Gandolfi:2008id, Gezerlis:2009iw,Vidana:2009is,
Piekarewicz:2009gb}.

While the overall moment of inertia of the neutron star is of great
interest, the {\sl fraction} of the moment of inertia contained in the
stellar crust may be as useful in constraining the equation of
state~\cite{Link:1999ca,Lattimer:2006xb}.  Indeed, by studying the
sudden and fairly regular spin jumps (``{\sl pulsar glitches}'') in
the Vela pulsar, it was determined that at least 1.4\% of the total
moment of inertia of the star must reside in the solid
crust~\cite{Link:1999ca}.  Pulsar glitches are believed to be the
result of angular momentum transfer between the star's solid crust and
a more rapidly rotating superfluid component residing in the stellar
interior.  As the crustal moment of inertia is sensitive to the
transition pressure at the core-crust interface~\cite{Link:1999ca},
the above ($\gtrsim$~1.4\%) limit may place a significant constraint
on the EOS of dilute, neutron-rich matter.  In this way, the crustal
moment of inertia may provide an attractive astrophysical complement
to PREx and to the dynamics of dilute Fermi gases in constraining the
density dependence of the symmetry energy.

The paper has been organized as follows. We start Sec.~\ref{Formalism}
by providing the background material necessary to compute the moment
of inertia of a neutron star in the {\sl slow-rotation}
approximation~\cite{Hartle:1967he,Hartle:1968si}. Later on in the
section we introduce the various accurately-calibrated relativistic
mean-field models that will be used to compute the equation of state
of the stellar material~\cite{Mueller:1996pm,
Lalazissis:1996rd,Lalazissis:1999,Todd-Rutel:2005fa}.  We note that
whereas all the models reproduce various experimentally measured
properties of finite nuclei, they differ significantly in their
predictions for the equation of state at both low and high
densities. In Sec.~\ref{Results} we present results for various
neutron-star properties and explore their possible correlation to the
neutron skin thickness of ${}^{208}$Pb. Finally,
Sec.~\ref{Conclusions} contains our concluding remarks.

\section{Formalism}
\label{Formalism}

In this section we develop the formalism required to compute 
the stellar moment of inertia. The section consists of two main
components: (a) the equations of stellar structure and (b) the 
neutron-star matter equation of state.

\subsection{Equations of stellar structure}

The cornerstone of our approach is the {\sl slow-rotation
approximation} pioneered by Hartle and
Thorne~\cite{Hartle:1967he,Hartle:1968si}. We
assume that the neutron star is rotating uniformly with 
a stellar frequency $\Omega$ that is far smaller than 
the Kepler frequency at the equator. That is,
\begin{equation}
  \Omega \ll \Omega_{\rm max} \approx \sqrt{\frac{GM}{R^3}} \ .
 \label{OmKepler}
\end{equation}
The formalism---even in the slow-rotation approximation---is subtle,
primarily due to the dragging of local inertial frames. The expression
for the moment of inertia of an axisymmetric star in hydrostatic
equilibrium is derived in great detail in
Refs.~\cite{Hartle:1973zza,Weber:1999}, so we only summarize here some
of the most important results. For a more pedagogical discussion one
may consult the text by Glendenning~\cite{Glendenning:2000}.

In the slow-rotation (first order in the angular velocity)
approximation the moment of inertia of a uniformly rotating, axially
symmetric neutron star is given by the following expression:
\begin{equation}
  I \equiv \frac{J}{\Omega} = \frac{8\pi}{3}
 \int_{0}^{R} r^{4} e^{-\nu(r)}\frac{\bar{\omega}(r)}{\Omega}
 \frac{\Big({\mathcal E}(r)+P(r)\Big)}{\sqrt{1-2GM(r)/r}} dr \;,
 \label{MomInertia}
\end{equation}
where $J$ is the angular momentum, $\nu(r)$ and $\bar{\omega}(r)$ are
radially-dependent metric functions (see below) and $M(r)$, ${\mathcal
E}(r)$, and $P(r)$ are the stellar mass, energy density, and pressure
profiles, respectively.  Perhaps the greatest advantage of the
slow-rotation approximation is that all the quantities appearing in
Eq.~(\ref{MomInertia}) may be assumed to remain spherically symmetric.
This implies that all stellar profiles may be determined from the 
Tolman-Oppenheimer-Volkoff (TOV) equation. That is,
\begin{subequations}
 \begin{align}
  & \frac{dP(r)}{dr} = -G \frac{\Big({\mathcal E}(r)+P(r)\Big)
      \Big(M(r)+4\pi r^{3}P(r)\Big)}{r^{2}\Big(1-2GM(r)/r\Big)} \;, \\
  & \frac{dM(r)}{dr} = 4\pi r^{2} {\mathcal E}(r) \;.
  \end{align}
 \label{TOV}
\end{subequations}
Given boundary conditions in terms of a central pressure
$P(0)\!=\!P_{c}$ and $M(0)\!=\!0$, the TOV equations may be solved
once an equation of state $(P\!=\!P({\mathcal E)})$ is supplied. In
particular, the stellar radius $R$ and mass $M$ are determined from
the following two conditions: $P(R)\!=\!0$ and $M\!=\!M(R)$.

Once mass and pressure profiles have been obtained, the full
space-time metric may be determined. We note that in the slow-rotation
approximation the invariant interval for the background metric of a
stationary, axially symmetric star may be written
as~\cite{Hartle:1967he,Landau:1975}:
\begin{equation}
 ds^2 = g_{\mu\nu}dx^{\mu}dx^{\nu} =
-e^{2\nu(r)}dt^2 + e^{2\lambda(r)} dr^2 +
  r^2 d\theta^2 + r^2 \sin^2\theta d\phi^2 - 
 2\omega(r)r^2\sin^2\theta dt d\phi \;.
 \label{metric}
\end{equation}
As alluded earlier, in the slow-rotation approximation the spherically
symmetric metric functions $\nu(r)$ and $\lambda(r)$ remain unchanged
from their values obtained in the limit of a non-rotating, static, and
spherically symmetric neutron star. In particular, $\lambda(r)$ is
simply related to the mass profile $M(r)$ by
\begin{equation}
  g_{11}(r)=e^{2\lambda(r)}=\Big(1-2GM(r)/r\Big)^{-1} \;.
 \label{grrmetric}
\end{equation}
Further, $\nu(r)$ can be determined from solving a first-order
differential equation or, equivalently, from evaluating the
following integral:
\begin{equation}
 \nu(r) = \frac{1}{2}\ln\left(1-\frac{2GM}{R}\right)
            -G\int_{r}^{R} \frac{\Big(M(x)+4\pi x^{3}P(x)\Big)}
              {x^{2}\Big(1-2GM(x)/x\Big)}dx \;.
\label{gttmetric}
\end{equation}
Finally, one must determine the metric function $\omega(r)$---the one
new ingredient that emerges from the slow rotation and which has no
counterpart in Newtonian gravity. The frequency $\omega(r)$ appears as
a consequence of the dragging of local inertial frames by the rotating
star; the so-called Lense-Thirring effect. The {\sl effective} (or relative)
frequency $\bar{\omega}(r)\!\equiv\!\Omega\!-\!\omega(r)$ appearing in
Eq.~(\ref{MomInertia}) represents the angular velocity of the fluid as
measured in a local inertial reference frame. In particular, the {\sl
dimensionless} relative frequency
$\widetilde{\omega}(r)\!\equiv\!\bar{\omega}(r)/\Omega$ satisfies the
following second-order differential equation:
\begin{equation}
 \frac{d}{dr}\left(r^{4}j(r)\frac{d\widetilde{\omega}(r)}{dr}\right)
 +4r^{3}\frac{dj(r)}{dr}\widetilde{\omega}(r) = 0\;,
 \label{OmegaBar}
\end{equation}
where 
\begin{equation}
  j(r)=e^{-\nu(r)-\lambda(r)} = 
  \begin{cases} 
    e^{-\nu(r)}\sqrt{1-2GM(r)/r}  & \text{if $r \le R\;,$}\\
    1 &\text{if $r > R\;.$}
  \end{cases}
\end{equation}
Note that $\widetilde{\omega}(r)$ is subject to the following two 
boundary conditions:
\begin{subequations}
 \begin{align}
  & \widetilde{\omega}'(0)=0 \;, 
  \label{BC1}\\
  & \widetilde{\omega}(R)+\frac{R}{3}\,\widetilde{\omega}'(R)=1\;.
  \label{BC2}
 \end{align}
\end{subequations}
Also note that in the slow-rotation approximation the moment of inertia 
does not depend on the stellar frequency $\Omega$. In practice, one 
chooses an arbitrary value for the central frequency
$\widetilde{\omega}_{c}\!=\!\widetilde{\omega}(0)$ and numerically
integrates Eq.~(\ref{OmegaBar}) up to the edge of the star. In
general, the boundary condition at the surface [Eq.~(\ref{BC2})] will
not be satisfied for an arbitrary choice of $\widetilde{\omega}_{c}$,
so one must rescale the function and its derivative by an appropriate 
constant to correct for the mismatch.

The procedure described above provides all the necessary steps to
compute the integrand in Eq.~(\ref{MomInertia}). The moment of inertia
is then obtained by performing the one remaining integral using
standard numerical techniques. Having solved for both
$\widetilde{\omega}(r)$ and $I$, one could check the consistency of
the formalism by ensuring that the following equation is satisfied:
\begin{equation}
  \widetilde{\omega}'(R) = \frac{6GI}{R^{4}} \;.
  \label{omegaR}
\end{equation}

\subsection{The crustal moment of inertia}

It has been suggested that {\sl pulsar glitches}, namely, the sudden
increase in the spin rate of pulsars, may place important constraints
on the equation of state~\cite{Link:1999ca, Lattimer:2000nx}. In
particular, an analysis based on a long time observation of glitches
of the Vela pulsar suggests that at least 1.4\% of the total moment of
inertia must reside in the non-uniform
crust~\cite{Link:1999ca,Lattimer:2000nx}. This is interesting as the
crustal moment of inertia is particularly sensitive to the transition
pressure at the core-crust interface and this observable is believed
to be correlated to the density-dependence of the symmetry
energy~\cite{Worley:2008cb,Xu:2008vz,Xu:2009vi,Moustakidis:2010zx}.
Thus, it is both interesting and enlightening to obtain analytic
expressions for the crustal moment of inertia.

The crustal moment of inertia is defined in terms of the general
expression provided in Eq.~(\ref{MomInertia}) but with the range
of the integral now limited from the transition (or core) radius 
$R_{t}$ to the stellar radius $R$. That is,
\begin{equation}
  I_{cr} = \frac{8\pi}{3}
 \int_{R_{t}}^{R} r^{4} e^{-\nu(r)}\widetilde{\omega}(r)
 \frac{\Big({\mathcal E}(r)+P(r)\Big)}{\sqrt{1-2GM(r)/r}} dr \;.
 \label{MomInertiaCr0}
\end{equation}
However, given that the crust is thin and the density within it is
low, several approximation have been proposed that help render 
the integral tractable~\cite{Lorenz:1992zz,Ravenhall:1994,
Link:1999ca,Lattimer:2000nx,Lattimer:2006xb}. The various 
approximations and details on how to evaluate the integral
are left to the appendix. In particular, we show that under those 
approximations the TOV equation may be solved exactly.  Here,
however, we simply quote our final result:
\begin{equation}
  I_{cr} \approx \frac{16\pi}{3}\frac{R_{t}^{6}P_{t}}{R_{s}}
 \left[1-\left(\frac{R_{s}}{R}\right)\left(\frac{I}{MR^{2}}\right)\right]
 \left[1+\frac{48}{5}(R_{t}/R_{s}-1)(P_{t}/{\mathcal E}_{t}) +
   \ldots\right] \;,
 \label{IcrFinal0}
\end{equation}
where $R_{s}\!=\!2GM$ is the Schwarzschild radius of the star, and
$P_{t}\!=\!P(R_{t})$ and ${\mathcal E}_{t}\!=\!{\mathcal E}(R_{t})$
are the pressure and energy density at the core-crust interface.  The
{\sl ellipsis} in the above equation indicates that the derivation
was carried out to first order in the small quantity $P_{t}/{\mathcal
E}_{t}\!\lesssim\!0.01$.  A few comments are in order. First, the
above expression for the crustal moment of inertia is extremely
accurate (of the order of a few percent; see Tables~\ref{Table4} 
and~\ref{Table5}). Second, the two terms in brackets in
Eq.~(\ref{IcrFinal0}) provide each a moderate $\sim$10\% correction to
the leading term, with the corrections being of opposite sign.  Third,
although the crustal moment of inertia still depends on the total
moment of inertia $I$, one may preserve the accuracy of our result
without having to rely on a highly accurate estimate of $I$. In
particular, by using the simple relationship proposed in
Ref.~\cite{Ravenhall:1994}, namely,
\begin{equation}
 \frac{I}{MR^{2}} = \frac{0.21}{1-R_{s}/R} \;,
 \label{RP94}
\end{equation}
one obtains
\begin{equation}
  I_{cr} \approx \frac{16\pi}{3}\frac{R_{t}^{6}P_{t}}{R_{s}}
 \left[1-\frac{0.21}{(R/R_{s}-1)}\right]
 \left[1+\frac{48}{5}(R_{t}/R_{s}-1)(P_{t}/{\mathcal E}_{t}) +
   \ldots\right] \;.
 \label{IcrFinal1}
\end{equation}
This expression remains extremely accurate, yet has the added appeal
that for a neutron star with a given mass $M$ (or $R_{s}$) and radius
$R$, the crustal moment of inertia depends exclusively on $R_{t}$,
$P_{t}$, and ${\mathcal E}_{t}$---all quantities that are expected to
be sensitive to the density dependence of the symmetry energy.  Note 
that the approximation for the total moment of inertia given in
Eq.~(\ref{RP94}) has been put into question in
Ref.~\cite{Lattimer:2000nx}. However, for our purposes such an
approximation is adequate as $I$ enters into the expression for the
crustal moment of inertia as a small correction. No such
approximation will be made for $I$ when reporting the {\sl fractional}
moment of inertia $I_{cr}/I$. Indeed, no approximation for $I$ will be
made at all. 

For completeness, we include an expression for the crustal mass
that was derived in the appendix following similar steps. We obtain,
\begin{equation}
  M_{cr} \approx 8\pi R_{t}^{3}P_{t}(R_{t}/R_{s}-1)
 \left[1+\frac{32}{5}(R_{t}/R_{s}-3/4)(P_{t}/{\mathcal E}_{t}) +
   \ldots\right] \;.
 \label{McrFinal1}
\end{equation}

\subsection{Neutron-star matter equation of state}

As alluded earlier, the structure of neutron stars is sensitive to the
equation of state of cold, fully catalyzed, neutron-rich matter.
Spanning many orders of magnitude in density, neutron stars display
rich and exotic phases that await a detailed theoretical
understanding.  For example, at densities that are about one third of
nuclear matter saturation density and below, the uniform ground state
becomes unstable against clustering correlations.  This non-uniform
region constitutes the stellar crust, which itself is divided into an
inner and an outer region.  In the outer crust the system is organized
into a Coulomb lattice of neutron-rich nuclei embedded in a degenerate
electron gas~\cite{Baym:1971pw,RocaMaza:2008ja}.  For this region we
employ the equation of state of Baym, Pethick, and
Sutherland~\cite{Baym:1971pw}. With increasing density the nuclei
become progressively more neutron rich until the neutron drip region
is reached; this region defines the boundary between the outer and the
inner crust.  It has been speculated that the bottom layers of the
inner crust consist of complex and exotic structures that are
collectively known as {\emph{nuclear
pasta}}~\cite{Ravenhall:1983uh,Hashimoto:1984,Lorenz:1992zz}.  Whereas
significant progress has been made in simulating this exotic
region~\cite{Horowitz:2004yf,Horowitz:2004pv,Horowitz:2005zb}, a
detailed equation of state is still missing. Hence, we resort to a
fairly accurate polytropic EOS to interpolate between the outer solid 
crust and the uniform liquid interior~\cite{Link:1999ca,Carriere:2002bx}.
For the uniform liquid core---with densities as low as one-third and
as high as ten times nuclear-matter saturation density---we generate
the equation of state using various refinements to the relativistic
mean-field model of Serot and
Walecka~\cite{Walecka:1974qa,Serot:1984ey,Serot:1997xg}.  For
consistency, the same models are used to compute the transition
density from the liquid core to the solid crust. This we do by
searching for the critical density at which the uniform system becomes
unstable to small amplitude density oscillations. The stability
analysis of the uniform ground state is based on a relativistic
random-phase-approximation (RPA) as detailed in
Ref.~\cite{Carriere:2002bx}.

The equation of state for the uniform liquid core is based on an
interacting Lagrangian that has been accurately calibrated to a
variety of ground-state properties of both finite nuclei and infinite 
nuclear matter. The model includes a nucleon field ($\psi$), two
isoscalar mesons (a scalar $\phi$ and a vector $V^{\mu}$), and one 
isovector meson ($b^{\mu}$)~\cite{Serot:1984ey,Serot:1997xg}
(the photon field plays no role in the present discussion of infinite 
nuclear matter at the mean-field level). The free Lagrangian density
for this model is given by
\begin{align}
{\mathscr L}_{0} & =
\bar\psi \left(i\gamma^{\mu}\partial_{\mu}\!-\!M\right)\psi +
\frac{1}{2}\partial_{\mu} \phi \partial^{\mu} \phi 
-\frac{1}{2}m_{s}^{2}\phi^{2} \nonumber \\
&-\frac{1}{4}F^{\mu\nu}F_{\mu\nu} +
\frac{1}{2}m_{v}^{2}V^{\mu}V_{\mu} - 
\frac{1}{4}{\bf b}^{\mu\nu}{\bf b}_{\mu\nu} +
\frac{1}{2}m_{\rho}^{2}{\bf b}^{\mu}{\bf b}_{\mu} \;,
\label{Lagrangian0}
\end{align}
where $F_{\mu\nu}$ and ${\bf b}_{\mu\nu}$ are the isoscalar
and isovector field tensors, respectively. That is,
\begin{subequations}
\begin{align}
 F_{\mu\nu} &= \partial_{\mu}V_{\nu} - \partial_{\nu}V_{\mu} \;, \\
 {\bf b}_{\mu\nu} &= \partial_{\mu}{\bf b}_{\nu} 
 - \partial_{\nu}{\bf b}_{\mu} \;.
\label{FieldTensors}
\end{align}
\end{subequations}
Further, the parameters $M$, $m_{s}$, $m_{v}$, and $m_{\rho}$ 
represent the nucleon and meson masses and may be treated 
(if wished) as empirical constants. The interacting Lagrangian
density is given by the following expression~\cite{Serot:1984ey,
Serot:1997xg,Mueller:1996pm}:
\begin{equation}
{\mathscr L}_{\rm int} =
\bar\psi \left[g_{\rm s}\phi   \!-\!
         \left(g_{\rm v}V_\mu  \!+\!
    \frac{g_{\rho}}{2}\mbox{\boldmath$\tau$}\cdot{\bf b}_{\mu}
          \right)\gamma^{\mu} \right]\psi -
          U(\phi,V^{\mu},{\bf b^{\mu}}) \;.
\label{Lagrangian}
\end{equation}
The model includes Yukawa couplings ($g_{\rm s}$, $g_{\rm v}$, and
$g_{\rho}$) between the nucleon and the three meson fields.  However,
to improve the phenomenological standing of the model---for example,
to soften the compressibility of symmetric nuclear matter---the
Lagrangian density must be supplemented by nonlinear meson
interactions. These are given by
\begin{equation}
  U(\phi,V^{\mu},{\bf b}^{\mu}) =
   \frac{\kappa}{3!} (g_{\rm s}\phi)^3 \!+\!
    \frac{\lambda}{4!}(g_{\rm s}\phi)^4  \!-\!
   \frac{\zeta}{4!}
    \Big(g_{\rm v}^2 V_{\mu}V^\mu\Big)^2 \!-\!
    \Lambda_{\rm v}
    \Big(g_{\rho}^{2}\,{\bf b}_{\mu}\cdot{\bf b}^{\mu}\Big)
    \Big(g_{\rm v}^2V_{\nu}V^\nu\Big) \;.
\label{USelf}
\end{equation}
Details on the calibration procedure may be found in
Refs.~\cite{Serot:1984ey,Serot:1997xg,Horowitz:2000xj,Todd:2003xs} 
and references therein. Note that additional local terms of the same order
in a power-counting scheme could have been included. However, their
phenomenological impact has been shown to be
small~\cite{Mueller:1996pm}, so they have not been included in the
calibration procedure.  Of particular interest and of critical
important to the present study is the {\sl isoscalar-isovector}
coupling term $\Lambda_{\rm v}$~\cite{Horowitz:2000xj,Todd:2003xs}.
Such a term has been added to the Lagrangian density to modify the
poorly known density dependence of the symmetry energy---a property
that is predicted to be stiff ({\sl i.e.,} to increase rapidly with
density) in most relativistic mean-field models. The addition of
$\Lambda_{\rm v}$ provides a simple---yet efficient and
reliable---method of tuning the density dependence of the symmetry
energy without sacrificing the success of the model in reproducing
experimentally constrained ground-state observables.  Because of the
sensitivity of the stellar radius to the density dependence of the
symmetry energy~\cite{Horowitz:2001ya}, we expect a strong correlation
between $\Lambda_{\rm v}$ and the stellar moment of inertia.

Whereas the full complexity of the quantum system can not be tackled,
the ground-state properties of the system may be computed in a {\sl
mean-field} (MF) approximation. In the MF approximation all the meson
fields are replaced by their classical expectation values and their
solution can be readily obtained by solving the classical
Euler-Lagrange equations of motion. The sole remnant of quantum
behavior is in the treatment of the nucleon field which emerges from 
a solution to the Dirac equation in the presence of appropriate
scalar and vector potentials~\cite{Serot:1984ey,Serot:1997xg}. 
Following standard mean-field practices, the energy density of 
the system is given by the following expression:
\begin{equation}
{\mathcal E}  = 
 \frac{1}{\pi^{2}}\int_{0}^{k_{\rm F}^{p}} k^{2}E_{k}^{\ast}\,dk +
 \frac{1}{\pi^{2}}\int_{0}^{k_{\rm F}^{n}} k^{2}E_{k}^{\ast}\,dk +
 g_{\rm v}V_{0}\rho_{\rm B} + \frac{g_{\rho}}{2}b_{0}\rho_{3} +
 U(\phi_{0},V_{0},b_{0}) \;,
\label{EDensity}
\end{equation}
where $E_{k}^{\ast}\!=\!\sqrt{k^{2}+M^{\ast 2}}$, 
$M^{\ast}\!=\!M-g_{\rm s}\phi_{0}$ is the effective nucleon mass, 
$k_{\rm F}^{p} (k_{\rm F}^{n})$ is the proton (neutron) Fermi momentum,
$\rho_{\rm B} (\rho_{3})$ is the isoscalar (isovector) baryon density,
and  $U(\phi_{0},V_{0},b_{0})$ is given by the following expression:
\begin{eqnarray}
  U(\phi_{0},V_{0},{\bf b}_{0}) & = &
   \frac{1}{2}m_{s}^{2}\phi_{0}^{2} +
   \frac{\kappa}{3!} (g_{\rm s}\phi_{0})^3 +
   \frac{\lambda}{4!}(g_{\rm s}\phi_{0})^4  \\ \nonumber & - &
   \frac{1}{2}m_{v}^{2}V_{0}^{2} -
   \frac{\zeta}{4!}(g_{\rm v}V_{0})^4 -
   \frac{1}{2}m_{\rho}^{2}b_{0}^{2} -
   \Lambda_{\rm v}(g_{\rm v}V_{0})^2 (g_{\rho}b_{0})^2 \;.  
 \label{USelfMFT}
\end{eqnarray}

The expression for the energy density may be ``simplified''
by using the classical equations of motion for the vector
fields to express the isoscalar and isovector densities 
$\rho_{\rm B}$ and $\rho_{3}$ in terms of $V_{0}$ and $b_{0}$.
One obtains, 
\begin{eqnarray}
{\mathcal E}  &=& 
 \frac{1}{\pi^{2}}\int_{0}^{k_{\rm F}^{p}} k^{2}E_{k}^{\ast}\,dk +
 \frac{1}{\pi^{2}}\int_{0}^{k_{\rm F}^{n}} k^{2}E_{k}^{\ast}\,dk 
 \nonumber \\
  &+& 
   \frac{1}{2}m_{s}^{2}\phi_{0}^{2} +
   \frac{\kappa}{3!} (g_{\rm s}\phi_{0})^3 +
   \frac{\lambda}{4!}(g_{\rm s}\phi_{0})^4 
 \nonumber \\
  &+& 
   \frac{1}{2}m_{v}^{2}V_{0}^{2} +
   \frac{\zeta}{8}(g_{\rm v}V_{0})^4 +
   \frac{1}{2}m_{\rho}^{2}b_{0}^{2} +
   3\Lambda_{\rm v}(g_{\rm v}V_{0})^2 (g_{\rho}b_{0})^2 \;.
\label{EDensity2}
\end{eqnarray}
Finally, as the MF approximation is {\sl thermodynamically
consistent}, the pressure of the system (at zero temperature) may be
obtained either from the energy-momentum tensor or from the energy
density and its first derivative with respect to the
density~\cite{Serot:1984ey,Serot:1997xg}.  That is,
\begin{equation}
 P  = 
 \frac{1}{3\pi^{2}}\int_{0}^{k_{\rm F}^{p}} \frac{k^{4}}{E_{k}^{\ast}}\,dk +
 \frac{1}{3\pi^{2}}\int_{0}^{k_{\rm F}^{n}} \frac{k^{4}}{E_{k}^{\ast}}\,dk -
U(\phi_{0},V_{0},b_{0}) \;,
\label{Pressure}
\end{equation}

Given that neutron-star matter is fully catalyzed, chemical
equilibrium must be imposed. Namely, the composition of the star is
determined from the equality of the chemical potential of the various
species. That is,
\begin{equation}
  \mu_{n} = \mu_{p} + \mu_{e} = \mu_{p} + \mu_{\mu} \;.
 \label{ChemicalEq}
\end{equation}
Note that electrons and muons are assumed to behave as 
relativistic free Fermi gases (with $m_{e}\!\equiv\!0$). Of 
course, muons appear in the system only after the electronic 
Fermi momentum becomes equal to the muon rest mass. The 
total energy density and pressure of the star are simply obtained 
by adding up the nucleonic and leptonic contributions.

\section{Results}
\label{Results}

\begin{figure}[tb]
\vspace{-0.05in}
\includegraphics[width=0.85\columnwidth,angle=0]{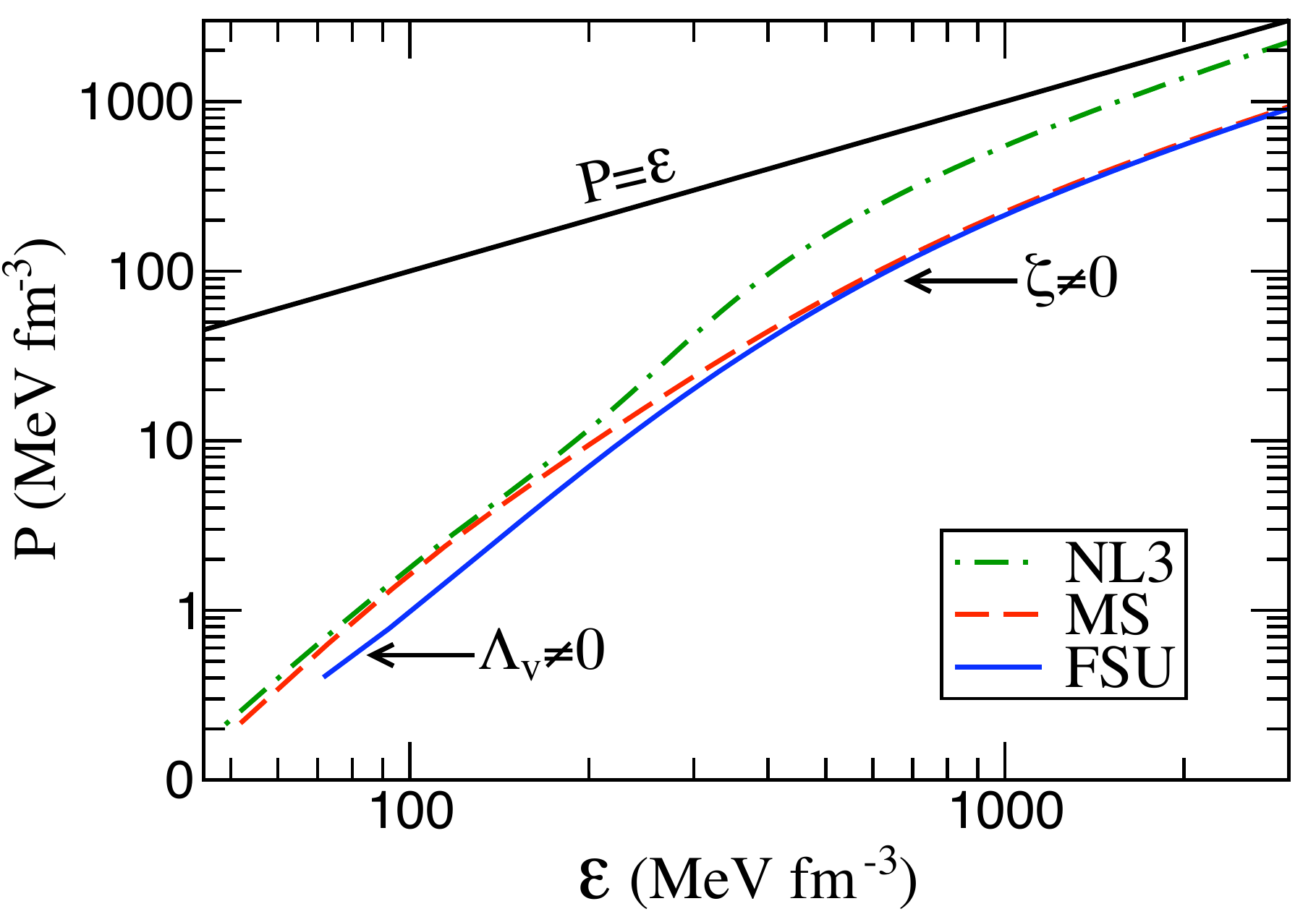}
\caption{(Color online) Equation of state ({\sl pressure vs
 energy density)} of neutron-star matter predicted by the 
 three relativistic mean-field models discussed in the text.
 The solid black line ($P\!=\!{\mathcal E}$) denotes the 
 stiffest possible equation of state consistent with causality.}
\label{Fig1}
\end{figure}

To study the sensitivity of the stellar moment of inertia to the
equation of state we will use relativistic mean-field models that have
been accurately calibrated to the properties of infinite-nuclear
matter (MS)~\cite{Mueller:1996pm}, to the ground-state properties of
finite nuclei (NL3)~\cite{Lalazissis:1996rd,Lalazissis:1999}, or to
both (FSUGold)~\cite{Todd-Rutel:2005fa}. Parameter sets for these
three models are listed in Table~\ref{Table1}.
\begin{widetext}
\begin{center}
\begin{table}[t]
\begin{tabular}{|l||c|c|c|c|c|c|c|c|c|c|}
 \hline
 Model & $m_{\rm s}$  & $m_{\rm v}$  & $m_{\rho}$  
       & $g_{\rm s}^2$ & $g_{\rm v}^2$ & $g_{\rho}^2$
       & $\kappa$ & $\lambda$ & $\zeta$ & $\Lambda_{\rm v}$\\
 \hline
 \hline
 NL3  & 508.194 & 782.501 & 763.000 & 104.3871 & 165.5854 &  79.6000 
         & 3.8599  & $-$0.01591 & 0.00 & 0.00 \\
 MS    & 485.000 & 782.500 & 763.000 & 111.0428 & 216.8998 &  70.5941 
          & 0.5083  & $+$0.02772 & 0.06 & 0.00 \\
 FSUGold   & 491.500 & 782.500 & 763.000 & 112.1996 & 204.5469 & 138.4701 
                  & 1.4203  & $+$0.02376 & 0.06 & 0.03 \\
\hline
\end{tabular}
\caption{Parameter sets (coupling constants and masses) for the
  mean-field models used in the text. The parameter $\kappa$ 
  and the meson masses $m_{\rm s}$, $m_{\rm v}$, and $m_{\rho}$ 
  are all given in MeV. The nucleon mass has been fixed at 
  $M\!=\!939$~MeV in all the models.}
\label{Table1}
\end{table}
\end{center}
\end{widetext}
The predicted equations of state---pressure {\sl vs} energy
density---for the three models are displayed in Fig.~\ref{Fig1}.
Given that the equation of state for the solid crust is identical in
all three models, we only present the contribution from the uniform
liquid core. The lowest energy density and pressure depicted in the
figure---which are different in all three models--- signal the
transition from the uniform liquid core to the non-uniform solid
crust.  Note that the uniform core is assumed to consist of nucleons
and leptons (electrons and muons) in chemical equilibrium; no exotic
degrees of freedom are considered.

The resulting equations of state show a significant model dependence.
As the models have been accurately calibrated, this is a clear
indication that the existing database of nuclear observables is
insensitive to both the low- and high-density behavior of the EOS (the
energy density at saturation is about $140~{\rm MeV/fm^{3}}$).  In
this contribution we are particularly interested in the sensitivity of
the stellar moment of inertia to the two empirical parameters $\zeta$
and $\Lambda_{\rm v}$---with the former controlling the high-density
behavior of the EOS and the latter the density dependence of the
symmetry energy.

\begin{figure}[tb]
\vspace{-0.05in}
\includegraphics[width=0.95\columnwidth,angle=0]{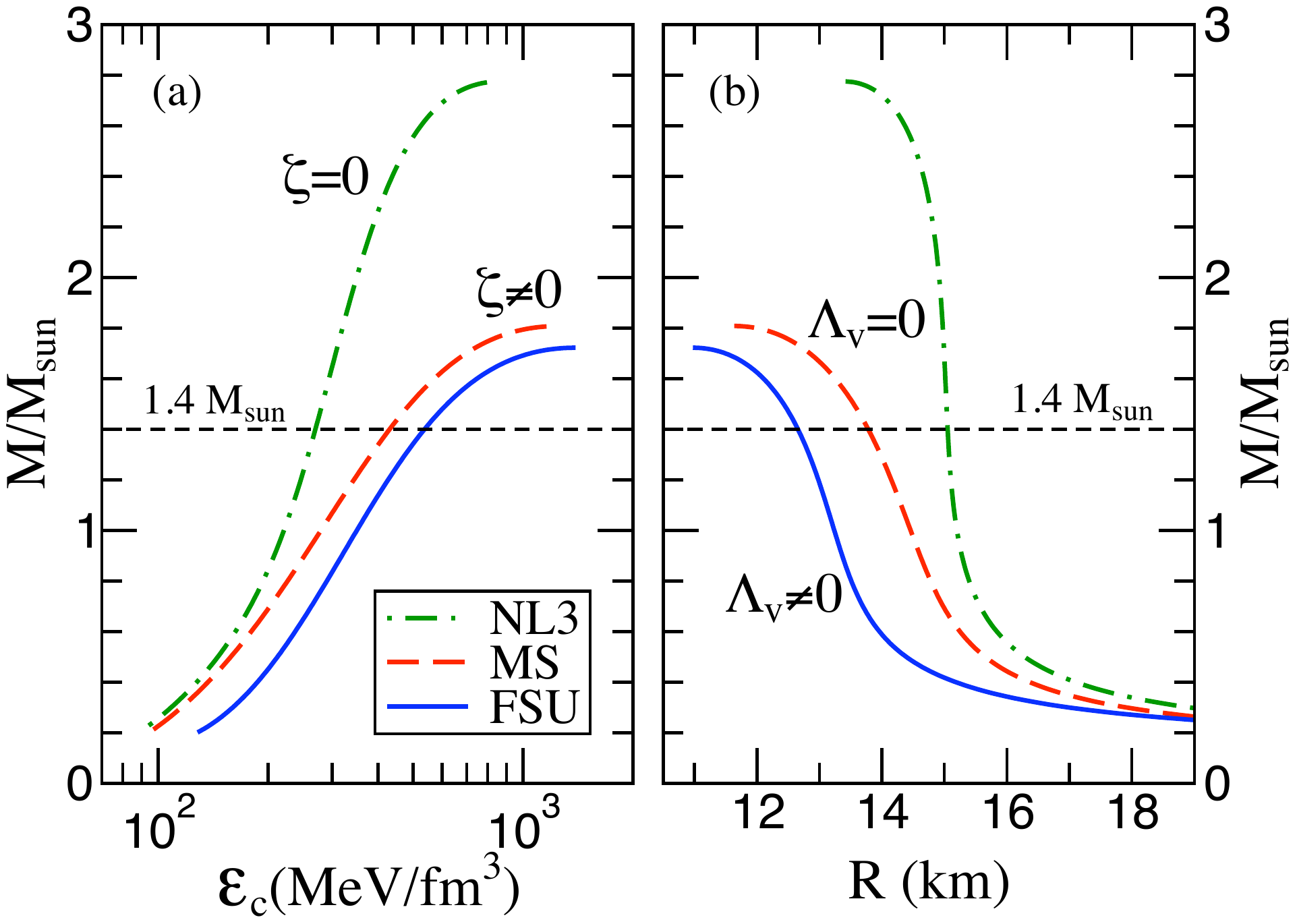}
\caption{(Color online) Neutron-star mass as a function of
  the central density (a) and the stellar radius (b) for the three 
  relativistic mean-field models discussed in the text.}
\label{Fig2}
\end{figure}

The NL3 parameter set~\cite{Lalazissis:1996rd,Lalazissis:1999}
provides an excellent description of the ground-state properties of
finite nuclei (such as masses and charge radii) without invoking
either $\zeta$ or $\Lambda_{\rm v}$ (see Table~\ref{Table1}).  As a
consequence of having set $\zeta\!=\!\Lambda_{\rm v}\!\equiv\!0$, 
NL3 generates a fairly stiff equation of state. At sub-saturation density,
this behavior is reflected in the relatively small value of the energy
density at the core-crust interface. At the high-density extreme, NL3
approaches the stiffest possible equation of state that is consistent
with causality ({\sl i.e.,} $P\!=\!{\mathcal E}$). As we shall see
below, such a stiff EOS generates neutron stars that are both 
massive and large (see Fig.~\ref{Fig2} and Table~\ref{Table2}).

As far as the MS equation of state is concerned, M\"uller and Serot
were able to build models with a wide range of values for $\zeta$ that
while reproducing the same observed properties at normal
nuclear densities, they produce maximum neutron star masses that
differ by almost one solar mass~\cite{Mueller:1996pm}.  By selecting a
value of $\zeta\!=\!0.06$, the MS model adopted here predicts a softer
EOS and consequently a limiting neutron-star mass that is
significantly smaller than NL3.  Such a softening at high density is
clearly evident in Fig.~\ref{Fig1}.  Note that the $\zeta\!=\!0.06$
choice appears consistent with the dense-matter equation of state
extracted from an analysis of energetic heavy-ion
collisions~\cite{Danielewicz:2002pu,Piekarewicz:2007dx}.  However,
extracting the {\sl zero-temperature} EOS from energetic heavy-ion
collisions may be model dependent.  Thus, observational data on
neutron-star masses may provide the cleanest constraint on the
high-density component of the equation of state. Note, however, that
since $\Lambda_{\rm v}$ remains equal to zero in this model, the
energy density and pressure at the core-crust interface remain largely
unchanged from their NL3 values (see Table~\ref{Table2}).

\begin{table}
\begin{tabular}{|l||c|c|c|c|c|c|}
  \hline
   Model & $M_{\rm max} (M_{\odot})$ & $R_{1.4}$~(km) 
       & ${\mathcal E}_{1.4}~({\rm MeV\,fm}^{-3})$  
       & $\rho_{t}~({\rm fm}^{-3})$ 
       & ${\mathcal E}_{t}~({\rm MeV\,fm}^{-3})$
       & $P_{t}~({\rm MeV\,fm}^{-3})$ 
       \\
  \hline
  \hline
   NL3           & 2.78 & 15.05  & 269.63 & 0.052 & 48.96 & 0.212 \\
   MS             & 1.81 & 13.78  & 430.81 & 0.055 & 51.91 & 0.216 \\
   FSUGold    & 1.72  & 12.66 & 536.91 & 0.076 & 71.53 & 0.402\\
  \hline
\end{tabular}
 \caption{Predictions for the maximum neutron star mass and for the 
   radius and central energy density of a 1.4~$M_{\odot}$ neutron star 
   in the three relativistic mean-field models discussed in the text. 
   The last three quantities represent the transition density, energy 
   density, and pressure at the core-crust interface.}
 \label{Table2}
\end{table}

The FSUGold parameter set is characterized by having both $\zeta$ and
$\Lambda_{\rm v}$ different from zero~\cite{Todd-Rutel:2005fa}.  By
adding $\Lambda_{\rm v}$ to the model one can soften the
EOS---particularly the symmetry energy---at low to intermediate
densities. This produces a shift of the core-crust transition energy
density and pressure to higher values relative to both NL3 and
MS. Moreover, such a softening generates neutron stars of relative
small radii (see Fig.~\ref{Fig2} and Table~\ref{Table2}). We note that
the softening of the symmetry energy is required to describe the
isoscalar monopole and isovector dipole modes in medium to heavy
nuclei~\cite{Piekarewicz:2003br}. Further, this softening appears
consistent with microscopic descriptions of the equation of state of
low-density neutron matter (see Ref.~\cite{Piekarewicz:2009gb} and
references therein).  On the other hand, the value of $\zeta\!=\!0.06$
adopted here is solely constrained by the dynamics of heavy ions. Yet,
the reported errors are large enough to accommodate slightly stiffer
equations of state (although not as stiff as
NL3)~\cite{Danielewicz:2002pu,Piekarewicz:2007dx}.

Having generated an equation of state, one can now proceed to solve 
the TOV equations [see Eqs.~(\ref{TOV})]. Once a value for the central
energy density ${\mathcal E}_{c}$ (or pressure $P_{c}$) is specified,
solutions to the TOV equations are obtained in the form of mass
$M(r)$, pressure $P(r)$, and energy density ${\mathcal E}(r)$ profiles.
From these, the stellar radius $R$ is extracted from the pressure 
profile as the point at which the pressure vanishes, namely, 
$P(R)\!=\!0$. Similarly, the total stellar mass is obtained from the
mass profile as $M\!=\!M(R)$. Note that for a given value of the
central energy density ${\mathcal E}_{c}$, a unique point in the 
$M$-$R$ diagram is generated.

In Fig.~\ref{Fig2} we display neutron-star masses as a function of the
central energy density (left panel) and the stellar radius (right
panel). The imprint of the underlying equation of state is clearly
evident on these curves. For example, the stiff behavior of the NL3
equation of state is reflected on its very large limiting mass (of
close to 3 solar masses). Also evident is the significant reduction in
the maximum stellar mass as one softens the EOS by shifting the value
of $\zeta$ from $0$ to $0.06$. Finally, we observe
a significant variation in the value of the central energy density
required to produce a ``canonical'' $1.4~M_{\odot}$ neutron star.  The
NL3 equation of state is so stiff ({\sl i.e.,} the pressure gradient
is so large) that a central energy density of only twice its value at
saturation is sufficient to support the star against gravitational
collapse. In contrast, the softer FSUGold equation of state requires 
a central energy density that is twice as large as NL3 or about 4 
times its value at saturation. 

Whereas $\zeta$ controls the maximum stellar mass, $\Lambda_{\rm v}$
controls the stellar radius. This is illustrated on the right-hand
panel of Fig.~\ref{Fig2}. Although the MS and FSUGold equations of
state display similar behavior at high density (see Fig.~\ref{Fig1})
the former generates stellar radii that are significantly larger than
the latter. For example, for a $1.4~M_{\odot}$ neutron star the
difference exceeds one kilometer.  Note that the stellar
radius---although primarily sensitive to the EOS at intermediate
densities---is also sensitive to the high-density component of the
EOS. Hence, although both NL3 and MS have $\Lambda_{\rm v}\!=\!0$, MS
(with $\zeta\!\ne\!0$) produces more compact stars.

\begin{widetext}
\begin{center}
\begin{table}
\begin{tabular}{|l||c|c|c|c|c|c|c|}
  \hline
   Model & $\Lambda_{\rm v}$ & $g_{\rho}^2$ & $L$~(MeV) & $R_{n}$-$R_{p}$~(fm)
             & $\rho_{t}~({\rm fm}^{-3})$  & ${\mathcal E}_{t}~({\rm MeV\,fm}^{-3})$
             & $P_{t}~({\rm MeV\,fm}^{-3})$ \\
  \hline
  \hline
   NL3    
   &  0.00 &  79.6000  & 118.189 & 0.280 & 0.052 & 48.960 & 0.212 \\
   &  0.01 &  90.9000  &  87.738  & 0.251 & 0.061 & 57.574 & 0.338 \\
   &  0.02 & 106.0000 &  68.217  & 0.223 & 0.746 & 70.630 & 0.508 \\
   &  0.03 & 127.1000 &  55.311  & 0.195 & 0.085 & 81.012 & 0.535 \\
   &  0.04 & 158.6000 &  46.607  & 0.166 & 0.090 & 85.618 & 0.376 \\
 \hline   
 FSU
   &  0.00 &  80.2618  & 108.764 & 0.286 & 0.051 & 48.458 & 0.207 \\    
   &  0.01 &  93.3409  &  87.276  & 0.260 & 0.060 & 56.330 & 0.317 \\
   &  0.02 & 111.5126 &  71.833  & 0.235 & 0.069 & 65.387 & 0.415 \\
   &  0.03 & 138.4701 &  60.515  & 0.207 & 0.076 & 71.534 & 0.402 \\
   &  0.04 & 182.6162 &  52.091  & 0.176 & 0.078 & 73.924 & 0.268 \\
   &  0.05 & 268.0859 &  45.743  & 0.137 & 0.077 & 73.206 & 0.036 \\
 \hline
\end{tabular}
 \caption{The NL3 and FSUGold {\sl ``families''} of mean-field interactions.
 The isovector parameters $\Lambda_{\rm v}$ and $g_{\rho}$ were adjusted so
 that all models have a symmetry energy of $\!\approx\!26$~MeV at
 a density of $\!\approx\!0.1~{\rm fm}^{-3}$. The tuning of the isovector 
 interaction modifies the slope of the symmetry energy $L$ at saturation 
 density. The impact of such a change on the neutron skin thickness of 
 ${}^{208}$Pb and on the transition density, energy density, and pressure 
 at the core-crust interface are displayed in the last four columns.}
 \label{Table3}
\end{table}
\end{center}
\end{widetext}
       
Having established the critical role of the two empirical parameters
$\zeta$ and $\Lambda_{\rm v}$ on the mass-vs-radius relationship of
neutron stars, we now return to the central goal of the present
manuscript: to assess the sensitivity of the stellar moment of
inertia---both total and crustal---to the equation of state. To
properly address this topic we must build models that, while
accurately calibrated, can generate a wide range of values for poorly
constrained nuclear observables. To do so, we modify the density
dependence of the symmetry energy by resorting to a simple---yet
highly robust---procedure first introduced in
Ref.~\cite{Horowitz:2000xj}. The procedure consists on modifying the
isovector mean-field interaction by simultaneously changing
$\Lambda_{\rm v}$ and $g_{\rho}$ (the $NN\rho$ coupling constant) in
such a way that the value of the symmetry energy remains fixed at a
specific value of the baryon density. Given that nuclei have a
low-density surface, the symmetry energy is best constrained 
not at nuclear matter saturation density, but at a slightly lower
value~\cite{Furnstahl:2001un}.  In this contribution---as in
Ref.~\cite{Horowitz:2000xj}---we fixed the symmetry energy at
$\approx\!26$~MeV at a density of $\approx\!0.1~{\rm fm}^{-3}$.  This
ensures that well constrained observables (such as masses and charge
radii) remain consistent with their experimental values.  Moreover, as
this procedure involves the tuning of only the isovector interaction,
all properties of symmetric nuclear matter remain intact.  Yet poorly
constrained observables---such as the neutron skin thickness of heavy
nuclei and neutron-star radii---can be made to vary over a wide range
of values~\cite{Horowitz:2000xj,Horowitz:2001ya}.  In
Table~\ref{Table3} we display the appropriate isovector parameters for
the NL3 and FSUGold {\sl ``families''} of mean-field interactions.
Particularly sensitive to this change is the slope of the symmetry
energy $L$ at saturation density. The table illustrates the impact of
$L$ on the neutron skin thickness of ${}^{208}$Pb and on the
transition density, energy density, and pressure at the core-crust 
interface. Note that there is no need to generate an MS family given
that it shares the same value of $\zeta$ with FSUGold.

\begin{figure}[tb]
\vspace{-0.05in}
\includegraphics[width=0.95\columnwidth,angle=0]{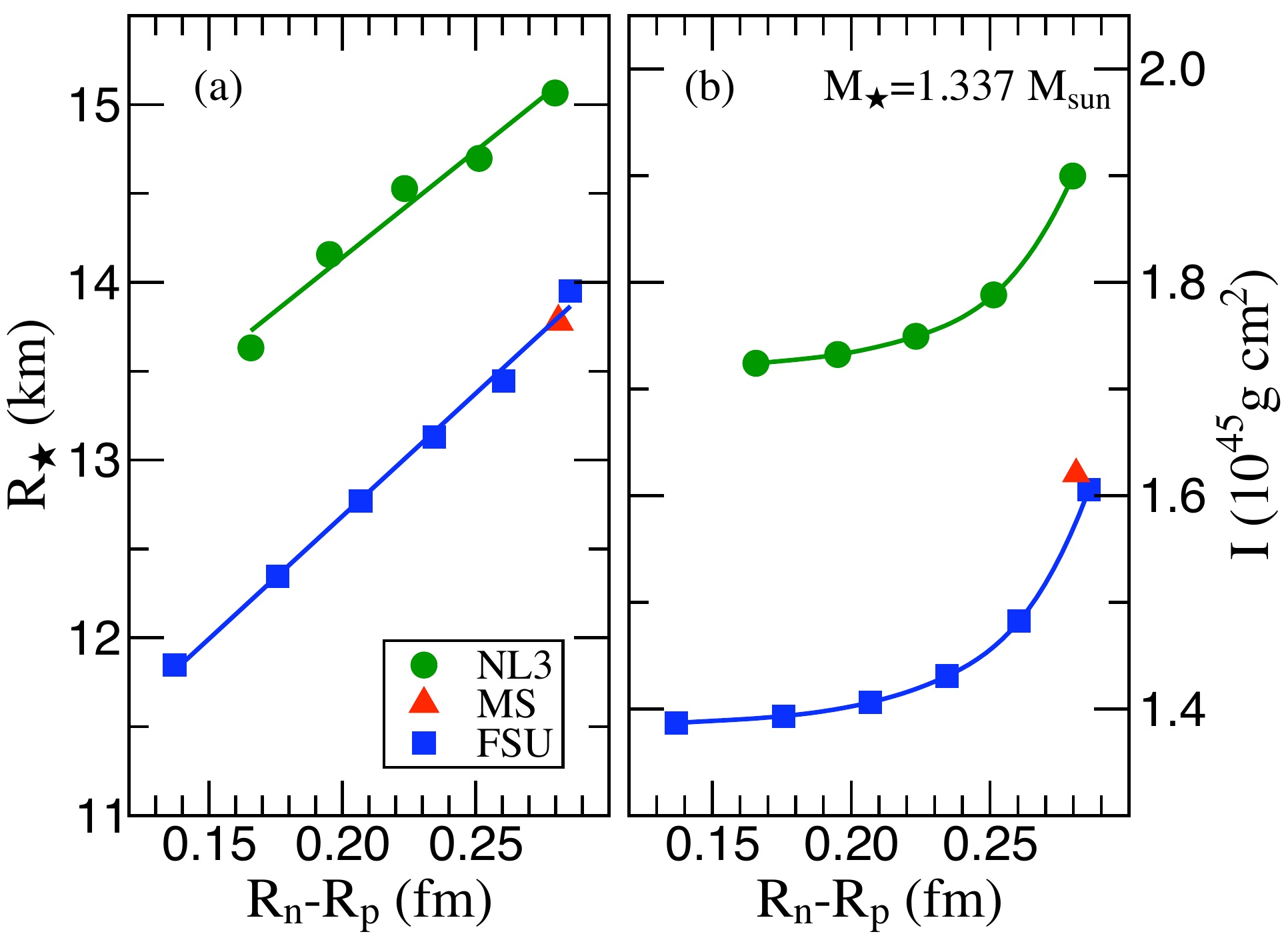}
\caption{(Color online) Stellar radius (a) and moment of inertia
  (b) of a $1.337~M_{\odot}$ neutron star (J0737-3039~A) as a 
  function of the neutron skin thickness of ${}^{208}$Pb.}
\label{Fig3}
\end{figure}

In Fig.~\ref{Fig3} we display the radius and moment of inertia of a
$M\!=\!1.337~M_{\odot}$ neutron star (such as J0737-3039~A) as a
function of the neutron skin thickness in ${}^{208}$Pb.  Although many
other observables (such as $L$) may be used to characterize the
density dependence of the symmetry energy, we have selected the
neutron-skin of ${}^{208}$Pb because it represents a fundamental
nuclear observable that will soon be directly determined from
laboratory data. The left-hand panel in the figure is reminiscent of
the linear correlation between the neutron skin in ${}^{208}$Pb and
the stellar radius uncovered in Ref.~\cite{Horowitz:2001ya}. Given
that neutron stars and the neutron skin of heavy nuclei are both made
of neutron-rich material, the emergence of such a correlation should
not come as a surprise. However, knowledge of the neutron skin is not
sufficient to determine the stellar radius. Whereas the neutron skin
of heavy nuclei is sensitive to the EOS around saturation density,
the stellar radius is also sensitive to its high-density component.
Hence, to eliminate the model dependence one must rely on
observational data rather than on laboratory experiments.  The
right-hand panel in Fig.~\ref{Fig3} displays the corresponding moment
of inertia. There is a significant drop in the moment of inertia as
the neutron skin departs from its largest value (at $\Lambda_{\rm
v}\!=\!0$) but then the sensitivity weakens. The same kind of behavior
is displayed when $I$ is plotted as a function of the stellar radius
(not shown). As in the case of the stellar radius, the total moment of
inertia is sensitive to the high-density component of the EOS, so a
strong model dependence remains. Note that since both MS and the 
stiffest member of the FSUGold family have $\Lambda_{\rm v}\!=\!0$ 
and $\zeta\!=\!0.06$, their predictions are very close to each other.

\begin{figure}[tb]
\vspace{-0.05in}
\includegraphics[width=0.95\columnwidth,angle=0]{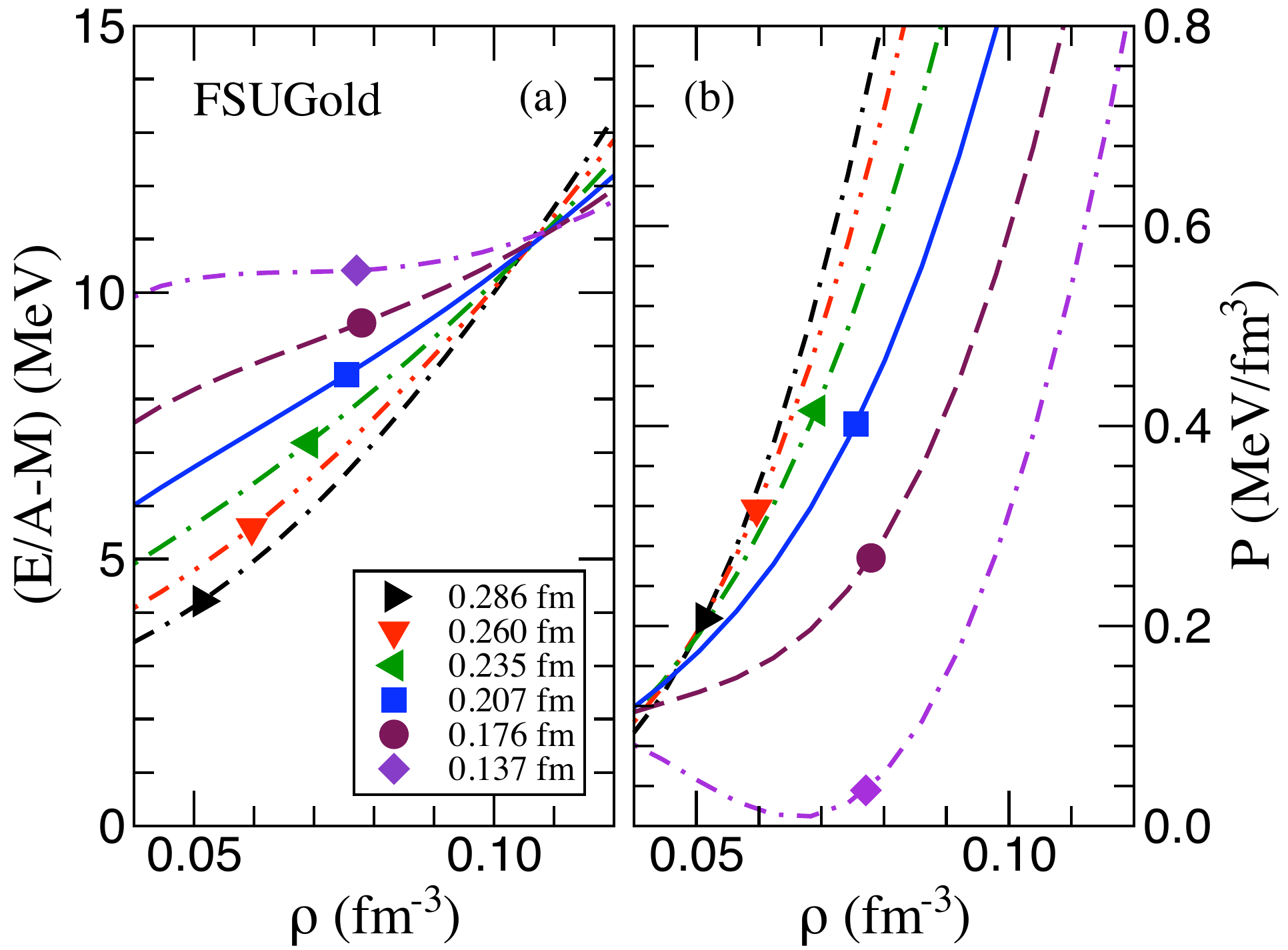}
\caption{(Color online) Equation of state for uniform, neutron-rich 
 matter in beta equilibrium for the FSUGold family of mean-field 
 interactions (labeled by the value of the neutron skin thickness in 
 ${}^{208}$Pb). The binding energy per nucleon (a) and the pressure 
 (b) are displayed in parametric form in terms of the baryon density.
 The various symbols indicate the transition density, energy
 per nucleon, and pressure at which the uniform state becomes 
 unstable against small density fluctuations.}
\label{Fig4}
\end{figure}

Given the sensitivity of the moment of inertia to the high density
component of the EOS, we now shift our attention to its crustal
component. Our expectation---based on its high sensitivity to the
transition pressure [see Eq.~(\ref{IcrFinal1})]---is that a strong
correlation will emerge between the crustal moment of inertia
($I_{cr}$) and the neutron skin thickness of ${}^{208}$Pb. To our
surprise, {\sl no such correlation exists}. We attempt to elucidate
this finding in what follows.

The core-crust boundary is determined by identifying the highest
baryon density at which the uniform ground state becomes unstable
against small amplitude density fluctuations.  The stability analysis
of the uniform ground state is based on a relativistic random-phase
approximation (RPA) that is described in detail in
Refs.~\cite{Horowitz:2000xj, Carriere:2002bx}. As first proposed in
Ref.~\cite{Horowitz:2000xj}---and confirmed since using various
equivalent approaches~\cite{Xu:2008vz,Xu:2009vi,Vidana:2009is,
Moustakidis:2010zx,Ducoin:2010as}---a strong correlations emerges
between the core-crust transition density ($\rho_{t}$) and the neutron
skin thickness of ${}^{208}$Pb. The left-hand panel in Fig.~\ref{Fig4}
provides evidence for such a correlation. The figure
displays the energy per nucleon as a function of baryon density for
uniform, neutron-rich matter in chemical equilibrium.  Results are
displayed for the various members of the FSUGold family of mean-field
interactions. The symbols in the figure are used to denote the
transition point and are labeled according to the value of
the neutron skin thickness in ${}^{208}$Pb. The behavior displayed in
Fig.~\ref{Fig4}(a) is simple to understand given the following facts:
(a) all models predict identical properties for symmetric nuclear
matter, (b) all models share the same value of the symmetry energy at
$\approx\!0.1~{\rm fm}^{-3}$, and (c) the softer the symmetry energy
the smaller the neutron skin thickness in ${}^{208}$Pb.  Based on
these fairly general assertions, one concludes that it is
energetically expensive for the system to remain uniform if the
equation of state is soft. Thus, models with thin neutron skins predict 
higher transition densities. 

However, whereas a clear correlation emerges between the transition
density and the neutron skin, {\sl no such correlations is observed in
the case of the transition pressure}. The right-hand panel in
Fig.~\ref{Fig4} displays the pressure of neutron-rich matter as a
function of the baryon density. As expected, 
the larger the neutron skin thickness in ${}^{208}$Pb the larger the
pressure.  That is, at a {\sl fixed given density}, the pressure is
larger for models that predict larger neutron skins.  However,
different models predict different transition densities and this mere
fact destroys the correlation between the transition pressure and the
neutron skin.  For example, for the three stiffest models displayed in
the figure, there is a direct correlation between the transition
density and the transition pressure ($P_{t}$). In this region these
three models exhibit relatively little scatter so that an increase in
$\rho_{t}$ is accompanied by a corresponding increase in
$P_{t}$. However, as the models continue to soften, an inverse
correlation develops: models with a soft symmetry energy have large
transition densities but small transition pressures. As a result, no
correlation between the transition pressure and the neutron skin
thickness in ${}^{208}$Pb develops. Note that this result is
consistent with the very recent analysis presented in
Ref.~\cite{Ducoin:2010as}.

\begin{figure}[htb]
\vspace{-0.05in}
\includegraphics[width=0.95\columnwidth,angle=0]{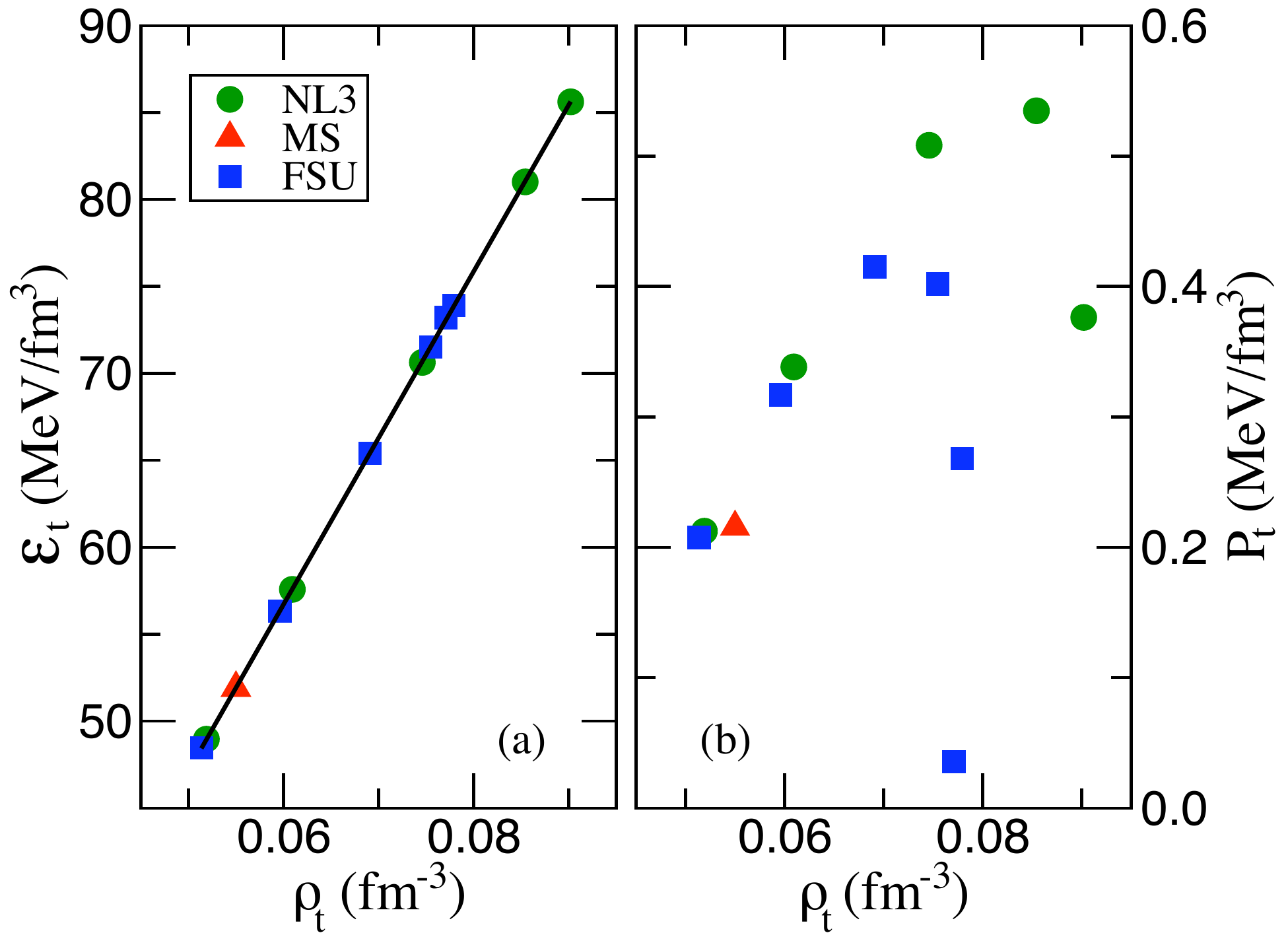}
\caption{(Color online) Properties of neutron-rich matter at the
 core-crust interface for the various models discussed in the
 text.  Whereas the transition energy density is strongly correlated 
 to the transition density (a), the transition pressure is not (b).}
\label{Fig5}
\end{figure}

\begin{figure}[htb]
\vspace{-0.05in}
\includegraphics[width=1.0\columnwidth,angle=0]{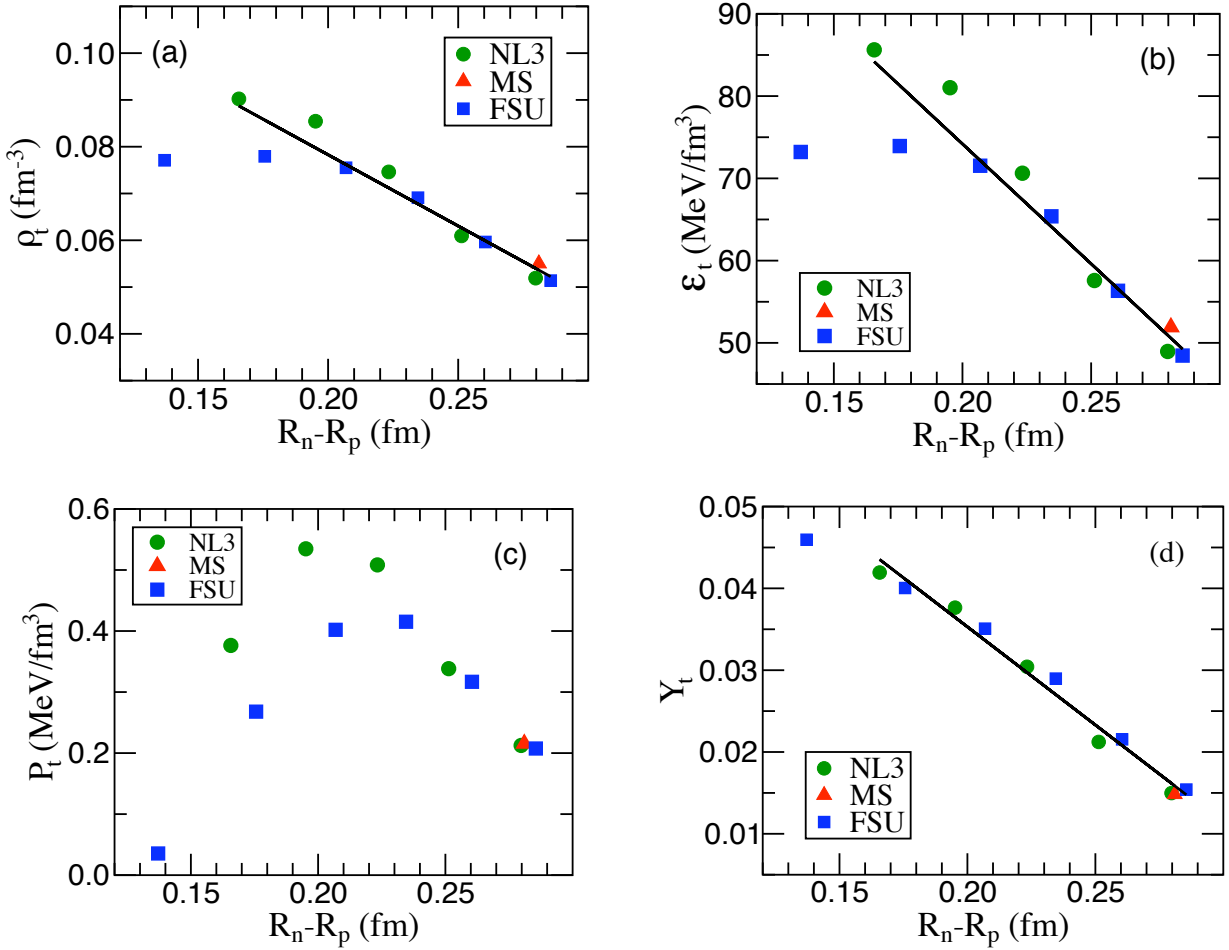}
\caption{(Color online) Baryon density (a), energy density (b),
pressure (c), and proton fraction (d) at the core-crust interface
as a function of the neutron skin thickness in ${}^{208}$Pb for
the various mean-field interactions discussed in the text.}
\label{Fig6}
\end{figure}

We now proceed to examine the correlation between the neutron skin
thickness of ${}^{208}$Pb and various quantities defined at the
transition region. Before doing so, however, we display in
Fig.~\ref{Fig5} the energy density and pressure at the core-crust
boundary as a function of the transition density. As suggested
earlier, whereas the energy density is strongly correlated to the
density at the interface, the pressure is not. In Fig.~\ref{Fig6} we
plot the value of various observables in the transition region as a
function of the neutron skin thickness in ${}^{208}$Pb.  In addition
to the transition density, energy density, and pressure, the proton
fraction is also displayed.  Excluding the softest model---which as we
shall see below appears in conflict with an observational
constraint---there is a clear (inverse) correlation between the
transition density $\rho_{t}$ and $R_{n}$-$R_{p}$, as originally
proposed in Ref.~\cite{Horowitz:2000xj}. Not surprisingly [see
Fig.~\ref{Fig5}(a)] the transition energy density displays an inverse
correlation that is just as strong. In contrast, there is no
correlation between the transition pressure and
$R_{n}$-$R_{p}$. Indeed, one can find models that have neutron skins
that vary by more than $0.1$~fm yet predict identical transition
pressures. Finally, the proton fraction at the core-crust interface
displays a tight inverse correlation. As mentioned earlier [see
Fig~\ref{Fig4}(a)] it is energetically expensive for models with a
soft-symmetry energy to support a large neutron-proton
asymmetry. Thus, the thinner the neutron skin of ${}^{208}$Pb, the
larger the proton fraction.

\begin{figure}[htb]
\vspace{-0.05in}
\includegraphics[width=0.9\columnwidth,angle=0]{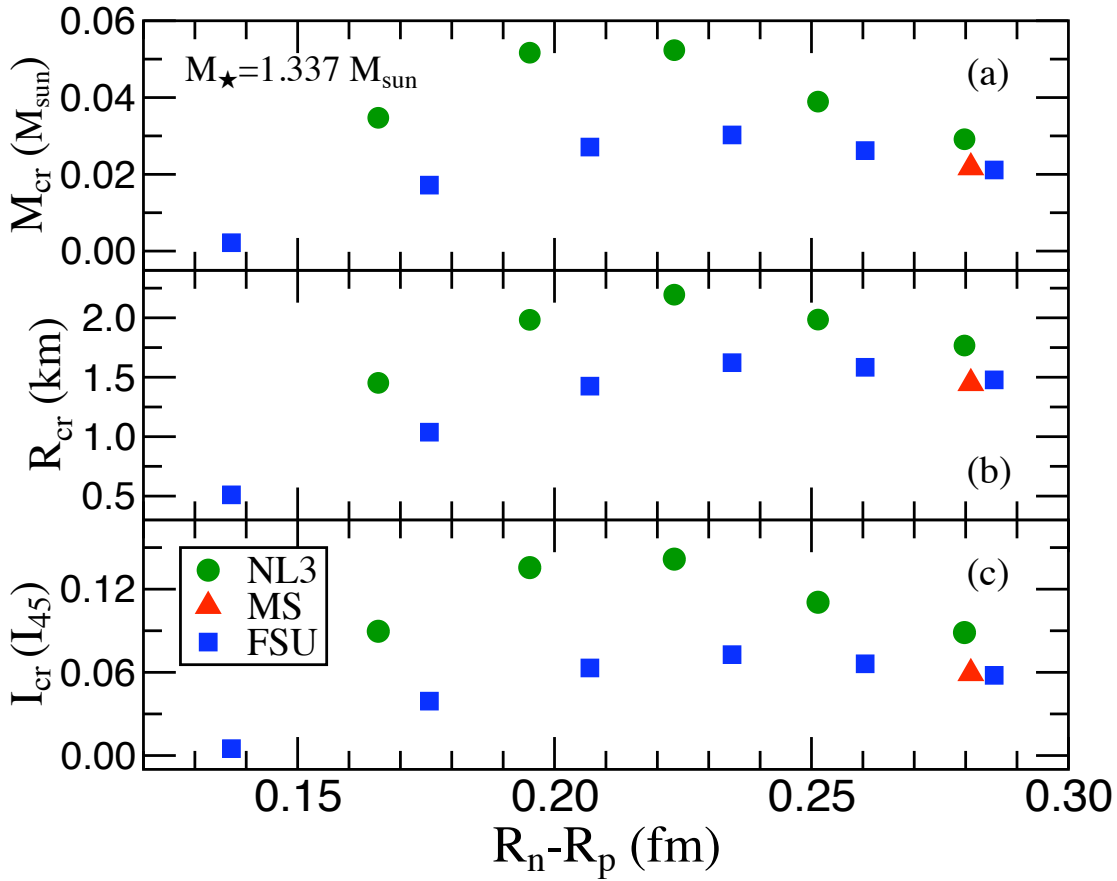}
\caption{(Color online) Crustal mass (a), crustal thickness (b), and
 crustal moment of inertia (c) as a function of the neutron skin 
thickness in ${}^{208}$Pb for a $1.337~M_{\odot}$ neutron star 
(J0737-3039~A). The various mean-field interactions are 
described in the text.}
\label{Fig7}
\end{figure}

Having explored the correlations (or lack-thereof) between the neutron
skin thickness in ${}^{208}$Pb and various properties of the EOS in
the transition region, we now proceed to study the sensitivity of
several crustal properties to $R_{n}$-$R_{p}$. In particular,
predictions for the crustal mass, thickness, and moment of inertia for
a $1.337~M_{\odot}$ neutron star (such as J0737-3039~A) are displayed
in Fig.~\ref{Fig7}. Note that approximate analytic expressions for the
crustal moment of inertia ($I_{cr}$) and mass ($M_{cr}$) have been
derived in the appendix and have been collected in
Eqs.~(\ref{IcrFinal1}) and~(\ref{McrFinal1}). Both of these
expressions indicate a high sensitivity to the transition pressure
$P_{t}$. Indeed, the imprint of the transition pressure is clearly
evident in all crustal observables.  For example, given that for
models with a soft symmetry energy the transition pressure increases
with $R_{n}$-$R_{p}$, an initial ({\sl i.e.,} for small neutron skins)
direct correlation develops between $R_{n}$-$R_{p}$ and all crustal
properties.  Eventually, however, the transition pressure reaches a
maximum and then an inverse correlation ensues. Hence, we conclude
that a measurement of the neutron skin thickness in ${}^{208}$Pb will
place no significant constraint on the crustal mass, thickness, or
moment of inertia. Note that the crustal thickness $R_{cr}$ also
follows such a trend due to its dependence on the crustal mass; if a
small crustal mass remains, then the crustal thickness will be small.

\begin{figure}[htb]
\vspace{-0.05in}
\includegraphics[width=0.9\columnwidth,angle=0]{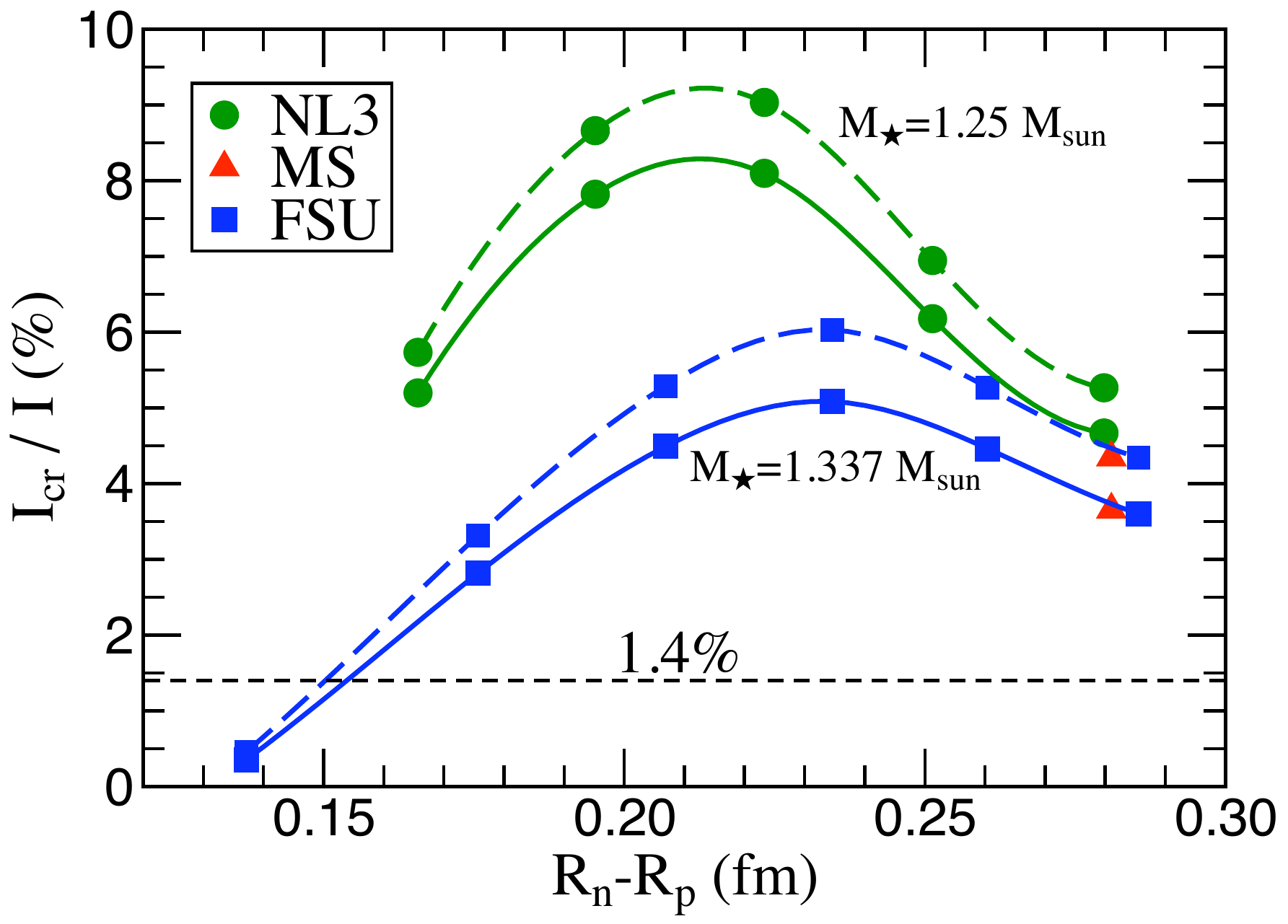}
\caption{(Color online)  Fraction of the crustal moment of inertia
 as a function of the neutron skin thickness in ${}^{208}$Pb for the
 binary pulsar J0737-3039 with masses of $1.337~M_{\odot}$ 
 (J0737-3039~A) and $1.25~M_{\odot}$  (J0737-3039~B). The 
 various mean-field interactions are described in the text.}
\label{Fig8}
\end{figure}

As mentioned in the Introduction, the study of pulsar glitches in the
Vela pulsar suggests that at least 1.4\% of the total moment of
inertia must reside in the solid
crust~\cite{Link:1999ca,Lattimer:2006xb}. The results displayed in
Fig.~\ref{Fig8} show how such a constrain may be used to rule out
certain equations of state. The figure shows predictions for the
fractional moment of inertia ($I_{cr}$) of the binary pulsar
J0737-3039 as a function of the neutron skin thickness in
${}^{208}$Pb. We observe that the softest member of the FSUGold family
predicts a fractional moment of inertia of only 0.35\%---significantly
lower than the 1.4\% bound. This suggests that models with such a soft
symmetry energy may be in conflict with observation.  Ultimately, such
a low value for $I_{cr}$ can be traced back to the very small
transition pressure predicted by the model. We close this section by
mentioning that many of the results presented and discussed in
graphical form have been collected in Tables~\ref{Table4}
and~\ref{Table5}.

\begin{widetext}
\begin{table}[ht]
\begin{tabular}{|l||c|c|c|c|c|c|c|c|c|}
\hline
 Model & $\Lambda_{\rm v}$ & $\rho_c$ & $R$  & $R_{core}$ & $M_{core}$
 & $M_{cr}/M (\%)$ & $I$ & $I_{core}$ & $I_{cr}/I (\%)$ \\ 
\hline \hline
 NL3 & 0.00 & 4.672 & 15.14 & 13.30 & 1.307 & 2.2(2.1) & 1.901 & 1.811 & 4.7(4.7) \\
         & 0.01 & 4.976 & 14.76 & 12.71 & 1.297 & 3.0(2.8) & 1.790 & 1.677 & 6.3(6.2) \\
         & 0.02 & 5.030 & 14.58 & 12.33 & 1.284 & 4.0(3.8) & 1.751 & 1.607 & 8.2(8.1) \\
         & 0.03 & 5.029 & 14.21 & 12.18 & 1.285 & 3.9(3.7) & 1.733 & 1.597 & 7.9(7.8) \\
         & 0.04 & 5.017 & 13.68 & 12.18 & 1.302 & 2.6(2.6) & 1.725 & 1.635 & 5.2(5.3) \\
\hline
 MS   & 0.00 & 7.136 & 13.92 & 12.45 & 1.315 & 1.6(1.6) & 1.620 & 1.560 & 3.7(3.7) \\
\hline
 FSU & 0.00 & 7.332 & 13.95 & 12.47 & 1.316 & 1.6(1.6) & 1.605 & 1.547 & 3.6(3.6) \\
        & 0.01 & 8.316 & 13.49 & 11.85 & 1.311 & 2.0(1.9) & 1.482 & 1.415 & 4.5(4.5) \\
        & 0.02 & 8.702 & 13.18 & 11.50 & 1.307 & 2.3(2.2) & 1.430 & 1.358 & 5.1(5.2) \\
        & 0.03 & 8.855 & 12.77 & 11.35 & 1.310 & 2.0(2.0) & 1.406 & 1.343 & 4.5(4.6) \\
        & 0.04 & 8.920 & 12.35 & 11.30 & 1.320 & 1.3(1.3) & 1.393 & 1.354 & 2.8(2.9) \\
        & 0.05 & 8.937 & 11.84 & 11.33 & 1.335 & 0.2(0.2) & 1.387 & 1.382 & 0.4(0.4) \\
\hline
\end{tabular}
\caption{Predictions for the properties of the J0737-3039 A pulsar
 with a mass of $M=1.337M_{\odot}$ and spin period of $P=22.7$~ms
 (or stellar frequency of $\Omega=276.8$ $s^{-1}$~\cite{Lyne:2004cj}).
 The central densities are in units of $10^{14} {\rm g/cm}^3$, the radii
 in km, the core mass in solar masses and the moments of inertia in
 $10^{45} {\rm g\,cm}^2$. The quantities in parenthesis are the analytic
 results for the fraction of the mass and moment of inertia contained
 in the solid crust (see text for details).}
\label{Table4}
\end{table}
\end{widetext}

\begin{widetext}
\begin{table}[ht]
\begin{tabular}{|l||c|c|c|c|c|c|c|c|c|}
\hline
 Model & $\Lambda_{\rm v}$ & $\rho_c$ & $R$  & $R_{core}$ & $M_{core}$
 & $M_{cr}/M (\%)$ & $I$ & $I_{core}$ & $I_{cr}/I (\%)$ \\ 
\hline \hline
 NL3 & 0.00 & 4.487 & 15.17 & 13.17 & 1.219 & 2.5(2.4) & 1.732 & 1.640 & 5.3(5.3) \\
         & 0.01 & 4.804 & 14.79 & 12.56 & 1.208 & 3.4(3.2) & 1.624 & 1.508 & 7.1(7.0) \\
         & 0.02 & 4.867 & 14.60 & 12.16 & 1.195 & 4.4(4.2) & 1.583 & 1.439 & 9.1(9.0) \\
         & 0.03 & 4.872 & 14.19 & 11.99 & 1.196 & 4.3(4.1) & 1.564 & 1.428 & 8.7(8.6) \\
         & 0.04 & 4.862 & 13.61 & 12.00 & 1.214 & 2.9(2.8) & 1.555 & 1.465 & 5.7(5.8) \\
\hline
 MS   & 0.00 & 6.474 & 14.08 & 12.44 & 1.226 & 1.9(1.9) & 1.507 & 1.442 & 4.3(4.3) \\
\hline
 FSU & 0.00 & 6.634 & 14.13 & 12.48 & 1.226 & 1.9(1.9) & 1.497 & 1.434 & 4.2(4.2) \\
        & 0.01 & 7.509 & 13.70 & 11.86 & 1.221 & 2.4(2.3) & 1.381 & 1.308 & 5.3(5.3) \\
        & 0.02 & 7.871 & 13.36 & 11.49 & 1.216 & 2.7(2.6) & 1.329 & 1.249 & 6.0(6.0) \\
        & 0.03 & 8.022 & 12.91 & 11.32 & 1.220 & 2.4(2.4) & 1.303 & 1.234 & 5.3(5.3) \\
        & 0.04 & 8.091 & 12.44 & 11.27 & 1.231 & 1.5(1.5) & 1.288 & 1.246 & 3.3(3.3) \\
        & 0.05 & 8.114 & 11.87 & 11.31 & 1.248 & 0.2(0.2) & 1.281 & 1.275 & 0.4(0.4) \\
\hline
\end{tabular}
\caption{Predictions for the properties of the J0737-3039 B pulsar
 with a mass of $M=1.250M_{\odot}$ and spin period of $P=2.77$~ms
 (or stellar frequency of $\Omega=2.268$ $s^{-1}$~\cite{Lyne:2004cj}).
 The central densities are in units of $10^{14} {\rm g/cm}^3$, the radii
 in km, the core mass in solar masses and the moments of inertia in
 $10^{45} {\rm g\,cm}^2$. The quantities in parenthesis are the analytic
 results for the fraction of the mass and moment of inertia contained
 in the solid crust (see text for details).}
\label{Table5}
\end{table}
\end{widetext}

\section{Conclusions}
\label{Conclusions}

In the present contribution we have studied the sensitivity of the
stellar moment of inertia to the underlying equation of state. In the
slow-rotation approximation employed here, the stellar moment of
inertia is only sensitive to the equation of state. Several equations
of state were generated using relativistic mean-field models that have
been accurately calibrated to the bulk properties of infinite nuclear
matter and finite nuclei.  As nuclear observables probe the EOS around
nuclear matter saturation density, two aspects of the EOS remain
poorly constrained even after the calibration procedure: (a) the
density dependence of the symmetry energy and (b) the high-density
component of the EOS. The relativistic mean-field models employed here
include two empirical parameters---$\Lambda_{\rm v}$ and
$\zeta$---that provide an efficient and economical way to modify the
EOS.  Whereas the former controls the density dependence of the
symmetry energy, the latter controls the high density component of the
EOS.  As such, $\Lambda_{\rm v}$ can be used to tune the pressure of
pure neutron matter at saturation density, thereby controlling the
neutron radius of objects as diverse as finite nuclei and neutron
stars. In contrast $\zeta$---by modifying the high-density component
of the EOS---strongly affects the maximum stellar mass that can be
supported against gravitational collapse.  By tuning these two
parameters, one can generate limiting stellar masses that differ by
more than one solar mass and radii (for a fixed stellar mass) that may
differ by more than 2~km. Note that one can generate these wide 
range of values without compromising the success of the models in 
reproducing a host of well determined nuclear observables.

With several equations of state in hand, we proceeded to compute the
moment of inertia of the recently discovered binary pulsar PSR
J0737-3039 (with individual masses of $M_{A}\!=\!1.337~M_{\odot}$ and
$M_{B}\!=\!1.250~M_{\odot}$). It has been suggested---due to the high
sensitivity of the binary to general relativistic effects---that the
moment of inertia of the fastest spinning pulsar in the binary (PSR
J0737-3039A) may be measured with a 10\% accuracy. Our results
indicate that knowledge of the pulsar moment of inertia (even with a
10\% accuracy) could help discriminate among various equations of
state.  We note, however, that whereas our results suggests that a
measurement of the moment of inertia could discriminate between
equations of state that are either stiff or soft at high density, the
sensitivity to the density dependence of the symmetry energy appears
to be weak---especially for models with a soft symmetry energy.

Although we find the total moment of inertia interesting, the main
focus of this contribution was its crustal component $I_{cr}$. Several
reasons prompted this choice. First, an analysis of pulsar glitches in
the Vela pulsar suggests that at least 1.4\% of the total moment of
inertia must reside in the solid crust. This places significant
constrains on the EOS.  Second, the crust is thin and the density
within it is low, so simple---yet fairly accurate---analytic
expressions for $I_{cr}$ exist. These indicate that the crustal moment
of inertia depends sensitively on a fundamental observable: the
transition pressure at the core-crust interface ($P_{t}$). Third,
given the strong correlation between the core-crust transition density
and the neutron skin thickness of ${}^{208}$Pb, one expects a
similar correlation to emerge in the case of $P_{t}$ and $I_{cr}$.
Finally, given that at the time of this writing the Parity Radius
Experiment (PREx) was actively collecting data, the prospects
of constraining crustal properties with laboratory data appears
imminent.

However, we found no correlation between the transition
pressure $P_{t}$ and the neutron skin thickness in ${}^{208}$Pb.
Whereas a robust correlation exists between $R_{n}$-$R_{p}$ and
various bulk properties of the EOS at the transition region---such as
the baryon density, energy density, and proton fraction---no such
correlation develops in the case of the transition pressure.  And
because of its sensitivity to the transition pressure, we conclude
that the crustal moment of inertia will not be significantly
constrained by a measurement of the neutron skin thickness in
${}^{208}$Pb. This represents the main conclusion of our work.  
We found the explanation for this behavior subtle and rooted on the
observed correlation between the transition density $\rho_{t}$
and $R_{n}$-$R_{p}$.  As expected, the larger the value of
$R_{n}$-$R_{p}$ the stiffer the symmetry energy. That is, given a
fixed value of the baryon density the resulting pressure increases
with $R_{n}$-$R_{p}$.  However, the transition pressure is obtained by
evaluating the EOS---not at a fixed value of the density, but
rather---at the {\sl transition density}, which is inversely
correlated to $R_{n}$-$R_{p}$.  It is precisely the fact that
$\rho_{t}$ changes with $R_{n}$-$R_{p}$ that destroys any correlation
between $P_{t}$ and $R_{n}$-$R_{p}$ [see Fig.~\ref{Fig4}(b)].
Although at odds with some studies that support the existence of such
a correlation~\cite{Worley:2008cb,Xu:2008vz,Xu:2009vi,
Moustakidis:2010zx}, our result appears consistent with a very recent
analysis by Ducoin, Margueron, and Providencia~\cite{Ducoin:2010as}.

\begin{acknowledgments}
 This work was supported in part by a grant from the U.S. 
 Department of Energy DE-FD05-92ER40750. 
\end{acknowledgments}

\appendix*
\section{}
In this appendix we describe the necessary steps used to obtain the
expression for the crustal moment of inertia given in
Eq.~(\ref{MomInertiaCr0}). Given that the uniform liquid core
accounts for most of the stellar mass and that the ``fluid'' in the
crust behave non-relativistically, the following three approximations
are assumed valid in the solid crust~\cite{Lattimer:2006xb}:
\begin{subequations}
 \begin{align}
  & M(r) \approx M(R)=M \;, \\
  & P(r) \ll {\mathcal E}(r) \;, \\
  & 4\pi r^{3}P(r) \ll M(r) = M \;.
 \end{align}
 \label{ThreeApproxs}
\end{subequations}
Under these assumptions the equations for stellar structure simplify
considerably. For example, Eq.~(\ref{gttmetric}) for the metric
$\nu(r)$ reduces to
\begin{equation}
 \nu(r) = \frac{1}{2}\ln\left(1-\frac{R_{s}}{R}\right)
            -\frac{1}{2}R_{s}\int_{r}^{R} \frac{dx}
              {x^{2}-xR_{s}} = \frac{1}{2}\ln\left(1-\frac{R_{s}}{r}\right)\;,
\label{NuofR}
\end{equation}
where $R_{s} \equiv 2GM$ is the Schwarzschild radius of the star.
Similarly, in the crustal region the TOV-equation [Eq.~(\ref{TOV})]
becomes equal to
\begin{equation}
 \frac{dP(r)}{dr} = -\frac{GM {\mathcal E}(r)}
  {r^{2}\Big(1-R_{s}/r\Big)} =-\frac{R_{s} {\mathcal E}(r)}
  {2r^{2}\Big(1-R_{s}/r\Big)}\;.
 \label{TOV2}
\end{equation}
If not for the (important) general-relativistic correction
$(1-R_{s}/r)^{-1}$, this expression would be identical to the equation
of hydrostatic equilibrium in the purely Newtonian limit.
Finally, the effective frequency $\bar\omega$ is approximated by its
value at $r\!=\!R$. That is,
\begin{equation}
 \frac{\bar{\omega}(r)}{\Omega} \approx
 \frac{\bar{\omega}(R)}{\Omega} =
 \left[1-\left(\frac{R_{s}}{R}\right)\left(\frac{I}{MR^{2}}\right)\right] \;,
 \label{LensThirring}
\end{equation}
where we have made use of  Eqs.~(\ref{BC2}) and~(\ref{omegaR}).

Using the above simplified expressions valid in the stellar crust, 
one obtains the following approximation for the crustal moment 
of  inertia~\cite{Link:1999ca,Lattimer:2000nx,Lattimer:2006xb}:
\begin{align}
  I_{cr} & = \frac{8\pi}{3}
 \left[1-\left(\frac{R_{s}}{R}\right)\left(\frac{I}{MR^{2}}\right)\right]
 \int_{R_{t}^{}}^{R} r^{4} \frac{{\mathcal E}(r)}{1-R_{s}/r} dr \nonumber\\
          & = \frac{16\pi}{3R_{s}}
 \left[1-\left(\frac{R_{s}}{R}\right)\left(\frac{I}{MR^{2}}\right)\right]
 \int_{0}^{P_{t}^{}} r^{6} dP \;.
 \label{Icrust}
\end{align}
where the last line follows from using the simplified TOV equation  
[see Eq.~(\ref{TOV2})]. To perform the above integral we need an
equation of state to compute the pressure profile $P\!=\!P(r)$ in the 
crust. As suggested in Refs.~\cite{Link:1999ca,Lattimer:2000nx}, a 
polytropic equation of state of index $\gamma\!=\!4/3$ will be
adopted for the crust. That is,
\begin{equation}
  P({\mathcal E})=K {\mathcal E}^{\gamma}=K {\mathcal E}^{4/3} \;,
 \label{Polytrope}
\end{equation}
where $K$ is a constant. 

Using such a simple---yet fairly accurate---EOS enables one to 
solve the TOV equation exactly. To do so, we first introduce the
following scaling variables:
\begin{equation}
 x \equiv r/R_{s} \;; \quad
 p \equiv P/P_{t} \;; \quad
 \epsilon \equiv {\mathcal E}/{\mathcal E}_{t}\;,
 \label{Scaling}
\end{equation}
where $P_{t}\!\equiv\!P(R_{t})$ and ${\mathcal E}_{t}\!\equiv\!{\mathcal E}(R_{t})$
are the pressure and energy density at the crust-core interface, with $R_{t}$ the 
transition (or core) radius. In terms of these scaling variables, the TOV-equation
in the crust [Eq.~(\ref{TOV2})] takes the following simple form:
\begin{equation}
  \xi\frac{dp}{dx}=-\frac{p^{1/\gamma}}{x(x-1)} \;,
 \label{TOV3}
\end{equation}
where the small parameter $\xi\!\equiv\!2P_{t}/{\mathcal E}_{t}$ (of the order of a
few percent) has been introduced. The above equation can now be integrated 
subject to the boundary condition $x\!=\!x_{t}\!=\!R_{t}/R_{s}$ at $p\!=\!1$. 
We obtain
\begin{align}
  x(p) &=\left(1-\left(1-x_{t}^{-1}\right)
 \exp\left[-\frac{\xi}{\alpha}\Big(p^{\alpha}-1\Big)\right]\right)^{-1} \\
         & \approx x_{t}\left[1+(1-x_{t})(p^{\alpha}-1)\frac{\xi}{\alpha}+
         \ldots\right]\;,
 \label{TOVSol}
\end{align}
where $\alpha\!=\!(\gamma-1)/\gamma\!=\!1/4$ and the second line provides
an approximation that is correct to first order in $\xi$. Although the integral
appearing in Eq.~(\ref{Icrust}) can now be performed using standard numerical
techniques, we prefer to provide an analytic expression for it by expanding the
integrand in powers of the small parameter $\xi$. And although the results 
presented here are correct only to first-order in $\xi$, analytic expressions can 
be developed to arbitrary order. We obtain for an arbitrary power $n$
\begin{equation}
  \int_{0}^{1}x^{n}(p) dp = x_{t}^{n}\left[1+\frac{n}{1+\alpha}(x_{t}-1)\xi + \ldots\right]
                                   = x_{t}^{n}\left[1+\frac{4n}{5}(x_{t}-1)\xi + \ldots\right] \;.
 \label{IcrInt}
\end{equation}
Substituting the above expression into Eq.~(\ref{Icrust}) we obtain the following 
analytic expression for the crustal moment of inertia to first order in 
$P_{t}/{\mathcal E}_{t}$:
\begin{equation}
  I_{cr} \approx \frac{16\pi}{3}\frac{R_{t}^{6}P_{t}}{R_{s}}
 \left[1-\left(\frac{R_{s}}{R}\right)\left(\frac{I}{MR^{2}}\right)\right]
 \left[1+\frac{48}{5}(R_{t}/R_{s}-1)(P_{t}/{\mathcal E}_{t}) + \ldots\right] \;.
\label{IcrApprox}
\end{equation}

Taking steps closely analogous to the ones followed for the crustal
moment of inertia, the fraction of the stellar mass contained in the 
solid crust may be written as
\begin{equation}
  M_{cr} = 4\pi \int_{R_{t}^{}}^{R} r^{2} {\mathcal E}(r) dr
      \approx 8\pi R_{s}^{3}P_{t}\int_{0}^{1}\Big(x^{4}(p)-x^{3}(p)\Big) dp \;.
 \label{Mcrust}
\end{equation}
The integral can now be easily performed with the aid of 
Eq.~(\ref{IcrInt}). We obtain
\begin{equation}
  M_{cr} \approx 8\pi R_{t}^{3}P_{t}(R_{t}/R_{s}-1)
 \left[1+\frac{32}{5}(R_{t}/R_{s}-3/4)(P_{t}/{\mathcal E}_{t}) +
   \ldots\right] \;.
 \label{Mcrust1}
\end{equation}

\vfill\eject
\bibliography{MomentInertia.bbl}

\begin{thebibliography}{58}
\expandafter\ifx\csname natexlab\endcsname\relax\def\natexlab#1{#1}\fi
\expandafter\ifx\csname bibnamefont\endcsname\relax
  \def\bibnamefont#1{#1}\fi
\expandafter\ifx\csname bibfnamefont\endcsname\relax
  \def\bibfnamefont#1{#1}\fi
\expandafter\ifx\csname citenamefont\endcsname\relax
  \def\citenamefont#1{#1}\fi
\expandafter\ifx\csname url\endcsname\relax
  \def\url#1{\texttt{#1}}\fi
\expandafter\ifx\csname urlprefix\endcsname\relax\def\urlprefix{URL }\fi
\providecommand{\bibinfo}[2]{#2}
\providecommand{\eprint}[2][]{\url{#2}}

\bibitem[{\citenamefont{Burgay et~al.}(2003)}]{Burgay:2003jj}
\bibinfo{author}{\bibfnamefont{M.}~\bibnamefont{Burgay}} \bibnamefont{et~al.},
  \bibinfo{journal}{Nature.} \textbf{\bibinfo{volume}{426}},
  \bibinfo{pages}{531} (\bibinfo{year}{2003}).

\bibitem[{\citenamefont{Lyne et~al.}(2004)}]{Lyne:2004cj}
\bibinfo{author}{\bibfnamefont{A.~G.} \bibnamefont{Lyne}} \bibnamefont{et~al.},
  \bibinfo{journal}{Science} \textbf{\bibinfo{volume}{303}},
  \bibinfo{pages}{1153} (\bibinfo{year}{2004}).

\bibitem[{\citenamefont{Hulse and Taylor}(1975)}]{Hulse:1974eb}
\bibinfo{author}{\bibfnamefont{R.~A.} \bibnamefont{Hulse}} \bibnamefont{and}
  \bibinfo{author}{\bibfnamefont{J.~H.} \bibnamefont{Taylor}},
  \bibinfo{journal}{Astrophys. J.} \textbf{\bibinfo{volume}{195}},
  \bibinfo{pages}{L51} (\bibinfo{year}{1975}).

\bibitem[{\citenamefont{Morrison et~al.}(2004)\citenamefont{Morrison,
  Baumgarte, Shapiro, and Pandharipande}}]{Morrison:2004df}
\bibinfo{author}{\bibfnamefont{I.~A.} \bibnamefont{Morrison}},
  \bibinfo{author}{\bibfnamefont{T.~W.} \bibnamefont{Baumgarte}},
  \bibinfo{author}{\bibfnamefont{S.~L.} \bibnamefont{Shapiro}},
  \bibnamefont{and} \bibinfo{author}{\bibfnamefont{V.~R.}
  \bibnamefont{Pandharipande}}, \bibinfo{journal}{Astrophys. J.}
  \textbf{\bibinfo{volume}{617}}, \bibinfo{pages}{L135} (\bibinfo{year}{2004}).

\bibitem[{\citenamefont{Lattimer and Schutz}(2005)}]{Lattimer:2004nj}
\bibinfo{author}{\bibfnamefont{J.~M.} \bibnamefont{Lattimer}} \bibnamefont{and}
  \bibinfo{author}{\bibfnamefont{B.~F.} \bibnamefont{Schutz}},
  \bibinfo{journal}{Astrophys. J.} \textbf{\bibinfo{volume}{629}},
  \bibinfo{pages}{979} (\bibinfo{year}{2005}).

\bibitem[{\citenamefont{Bejger et~al.}(2005)\citenamefont{Bejger, Bulik, and
  Haensel}}]{Bejger:2005jy}
\bibinfo{author}{\bibfnamefont{M.}~\bibnamefont{Bejger}},
  \bibinfo{author}{\bibfnamefont{T.}~\bibnamefont{Bulik}}, \bibnamefont{and}
  \bibinfo{author}{\bibfnamefont{P.}~\bibnamefont{Haensel}},
  \bibinfo{journal}{Mon. Not. Roy. Astron. Soc.}
  \textbf{\bibinfo{volume}{364}}, \bibinfo{pages}{635} (\bibinfo{year}{2005}).

\bibitem[{\citenamefont{Lavagetto et~al.}(2006)\citenamefont{Lavagetto,
  Bombaci, D'Ai', Vidana, and Robba}}]{Lavagetto:2006ew}
\bibinfo{author}{\bibfnamefont{G.}~\bibnamefont{Lavagetto}},
  \bibinfo{author}{\bibfnamefont{I.}~\bibnamefont{Bombaci}},
  \bibinfo{author}{\bibfnamefont{A.}~\bibnamefont{D'Ai'}},
  \bibinfo{author}{\bibfnamefont{I.}~\bibnamefont{Vidana}}, \bibnamefont{and}
  \bibinfo{author}{\bibfnamefont{N.~R.} \bibnamefont{Robba}}
  (\bibinfo{year}{2006}), \eprint{astro-ph/0612061}.

\bibitem[{\citenamefont{Iorio}(2008)}]{Iorio:2007yz}
\bibinfo{author}{\bibfnamefont{L.}~\bibnamefont{Iorio}}, \bibinfo{journal}{New
  Astron.} \textbf{\bibinfo{volume}{14}}, \bibinfo{pages}{40}
  (\bibinfo{year}{2008}).

\bibitem[{\citenamefont{Horowitz and
  Piekarewicz}(2001{\natexlab{a}})}]{Horowitz:2001ya}
\bibinfo{author}{\bibfnamefont{C.~J.} \bibnamefont{Horowitz}} \bibnamefont{and}
  \bibinfo{author}{\bibfnamefont{J.}~\bibnamefont{Piekarewicz}},
  \bibinfo{journal}{Phys. Rev.} \textbf{\bibinfo{volume}{C64}},
  \bibinfo{pages}{062802} (\bibinfo{year}{2001}{\natexlab{a}}).

\bibitem[{\citenamefont{Horowitz et~al.}(2001)\citenamefont{Horowitz, Pollock,
  Souder, and Michaels}}]{Horowitz:1999fk}
\bibinfo{author}{\bibfnamefont{C.~J.} \bibnamefont{Horowitz}},
  \bibinfo{author}{\bibfnamefont{S.~J.} \bibnamefont{Pollock}},
  \bibinfo{author}{\bibfnamefont{P.~A.} \bibnamefont{Souder}},
  \bibnamefont{and} \bibinfo{author}{\bibfnamefont{R.}~\bibnamefont{Michaels}},
  \bibinfo{journal}{Phys. Rev.} \textbf{\bibinfo{volume}{C63}},
  \bibinfo{pages}{025501} (\bibinfo{year}{2001}).

\bibitem[{\citenamefont{Kumar et~al.}(2005)\citenamefont{Kumar, Michaels,
  Souder, and Urciuoli}}]{Michaels:2005}
\bibinfo{author}{\bibfnamefont{K.}~\bibnamefont{Kumar}},
  \bibinfo{author}{\bibfnamefont{R.}~\bibnamefont{Michaels}},
  \bibinfo{author}{\bibfnamefont{P.~A.} \bibnamefont{Souder}},
  \bibnamefont{and} \bibinfo{author}{\bibfnamefont{G.~M.}
  \bibnamefont{Urciuoli}} (\bibinfo{year}{2005}),
  \urlprefix\url{http://hallaweb.jlab.org/parity/prex}.

\bibitem[{\citenamefont{Baker}(1999)}]{Baker:1999dg}
\bibinfo{author}{\bibfnamefont{G.~A.} \bibnamefont{Baker}},
  \bibinfo{journal}{Phys. Rev.} \textbf{\bibinfo{volume}{C60}},
  \bibinfo{pages}{054311} (\bibinfo{year}{1999}).

\bibitem[{\citenamefont{Heiselberg}(2002)}]{Heiselberg:2000bm}
\bibinfo{author}{\bibfnamefont{H.}~\bibnamefont{Heiselberg}},
  \bibinfo{journal}{Phys. Rev.} \textbf{\bibinfo{volume}{A63}},
  \bibinfo{pages}{043606} (\bibinfo{year}{2002}).

\bibitem[{\citenamefont{Carlson et~al.}(2003)\citenamefont{Carlson, Chang,
  Pandharipande, and Schmidt}}]{Carlson:2003}
\bibinfo{author}{\bibfnamefont{J.}~\bibnamefont{Carlson}},
  \bibinfo{author}{\bibfnamefont{S.-Y.} \bibnamefont{Chang}},
  \bibinfo{author}{\bibfnamefont{V.~R.} \bibnamefont{Pandharipande}},
  \bibnamefont{and} \bibinfo{author}{\bibfnamefont{K.~E.}
  \bibnamefont{Schmidt}}, \bibinfo{journal}{Phys. Rev. Lett.}
  \textbf{\bibinfo{volume}{91}}, \bibinfo{pages}{050401}
  (\bibinfo{year}{2003}).

\bibitem[{\citenamefont{Schwenk and Pethick}(2005)}]{Schwenk:2005ka}
\bibinfo{author}{\bibfnamefont{A.}~\bibnamefont{Schwenk}} \bibnamefont{and}
  \bibinfo{author}{\bibfnamefont{C.~J.} \bibnamefont{Pethick}},
  \bibinfo{journal}{Phys. Rev. Lett.} \textbf{\bibinfo{volume}{95}},
  \bibinfo{pages}{160401} (\bibinfo{year}{2005}).

\bibitem[{\citenamefont{Nishida and Son}(2006)}]{Nishida:2006br}
\bibinfo{author}{\bibfnamefont{Y.}~\bibnamefont{Nishida}} \bibnamefont{and}
  \bibinfo{author}{\bibfnamefont{D.~T.} \bibnamefont{Son}},
  \bibinfo{journal}{Phys. Rev. Lett.} \textbf{\bibinfo{volume}{97}},
  \bibinfo{pages}{050403} (\bibinfo{year}{2006}).

\bibitem[{\citenamefont{Hebeler and Schwenk}(2009)}]{Hebeler:2009iv}
\bibinfo{author}{\bibfnamefont{K.}~\bibnamefont{Hebeler}} \bibnamefont{and}
  \bibinfo{author}{\bibfnamefont{A.}~\bibnamefont{Schwenk}}
  (\bibinfo{year}{2009}), \eprint{0911.0483}.

\bibitem[{\citenamefont{Gandolfi et~al.}(2008)\citenamefont{Gandolfi,
  Illarionov, Fantoni, Pederiva, and Schmidt}}]{Gandolfi:2008id}
\bibinfo{author}{\bibfnamefont{S.}~\bibnamefont{Gandolfi}},
  \bibinfo{author}{\bibfnamefont{A.~Y.} \bibnamefont{Illarionov}},
  \bibinfo{author}{\bibfnamefont{S.}~\bibnamefont{Fantoni}},
  \bibinfo{author}{\bibfnamefont{F.}~\bibnamefont{Pederiva}}, \bibnamefont{and}
  \bibinfo{author}{\bibfnamefont{K.~E.} \bibnamefont{Schmidt}},
  \bibinfo{journal}{Phys. Rev. Lett.} \textbf{\bibinfo{volume}{101}},
  \bibinfo{pages}{132501} (\bibinfo{year}{2008}).

\bibitem[{\citenamefont{Gezerlis and Carlson}(2009)}]{Gezerlis:2009iw}
\bibinfo{author}{\bibfnamefont{A.}~\bibnamefont{Gezerlis}} \bibnamefont{and}
  \bibinfo{author}{\bibfnamefont{J.}~\bibnamefont{Carlson}}
  (\bibinfo{year}{2009}), \eprint{0911.3907}.

\bibitem[{\citenamefont{Piekarewicz}(2010)}]{Piekarewicz:2009gb}
\bibinfo{author}{\bibfnamefont{J.}~\bibnamefont{Piekarewicz}},
  \bibinfo{journal}{J. Phys.} \textbf{\bibinfo{volume}{G37}},
  \bibinfo{pages}{064038} (\bibinfo{year}{2010}), \eprint{0912.5103}.

\bibitem[{\citenamefont{Vidana et~al.}(2009)\citenamefont{Vidana, Providencia,
  Polls, and Rios}}]{Vidana:2009is}
\bibinfo{author}{\bibfnamefont{I.}~\bibnamefont{Vidana}},
  \bibinfo{author}{\bibfnamefont{C.}~\bibnamefont{Providencia}},
  \bibinfo{author}{\bibfnamefont{A.}~\bibnamefont{Polls}}, \bibnamefont{and}
  \bibinfo{author}{\bibfnamefont{A.}~\bibnamefont{Rios}},
  \bibinfo{journal}{Phys. Rev.} \textbf{\bibinfo{volume}{C80}},
  \bibinfo{pages}{045806} (\bibinfo{year}{2009}).

\bibitem[{\citenamefont{Link et~al.}(1999)\citenamefont{Link, Epstein, and
  Lattimer}}]{Link:1999ca}
\bibinfo{author}{\bibfnamefont{B.}~\bibnamefont{Link}},
  \bibinfo{author}{\bibfnamefont{R.~I.} \bibnamefont{Epstein}},
  \bibnamefont{and} \bibinfo{author}{\bibfnamefont{J.~M.}
  \bibnamefont{Lattimer}}, \bibinfo{journal}{Phys. Rev. Lett.}
  \textbf{\bibinfo{volume}{83}}, \bibinfo{pages}{3362} (\bibinfo{year}{1999}).

\bibitem[{\citenamefont{Lattimer and Prakash}(2007)}]{Lattimer:2006xb}
\bibinfo{author}{\bibfnamefont{J.~M.} \bibnamefont{Lattimer}} \bibnamefont{and}
  \bibinfo{author}{\bibfnamefont{M.}~\bibnamefont{Prakash}},
  \bibinfo{journal}{Phys. Rept.} \textbf{\bibinfo{volume}{442}},
  \bibinfo{pages}{109} (\bibinfo{year}{2007}).

\bibitem[{\citenamefont{Hartle}(1967)}]{Hartle:1967he}
\bibinfo{author}{\bibfnamefont{J.~B.} \bibnamefont{Hartle}},
  \bibinfo{journal}{Astrophys. J.} \textbf{\bibinfo{volume}{150}},
  \bibinfo{pages}{1005} (\bibinfo{year}{1967}).

\bibitem[{\citenamefont{Hartle and Thorne}(1968)}]{Hartle:1968si}
\bibinfo{author}{\bibfnamefont{J.~B.} \bibnamefont{Hartle}} \bibnamefont{and}
  \bibinfo{author}{\bibfnamefont{K.~S.} \bibnamefont{Thorne}},
  \bibinfo{journal}{Astrophys. J.} \textbf{\bibinfo{volume}{153}},
  \bibinfo{pages}{807} (\bibinfo{year}{1968}).

\bibitem[{\citenamefont{Mueller and Serot}(1996)}]{Mueller:1996pm}
\bibinfo{author}{\bibfnamefont{H.}~\bibnamefont{Mueller}} \bibnamefont{and}
  \bibinfo{author}{\bibfnamefont{B.~D.} \bibnamefont{Serot}},
  \bibinfo{journal}{Nucl. Phys.} \textbf{\bibinfo{volume}{A606}},
  \bibinfo{pages}{508} (\bibinfo{year}{1996}).

\bibitem[{\citenamefont{Lalazissis et~al.}(1997)\citenamefont{Lalazissis,
  Konig, and Ring}}]{Lalazissis:1996rd}
\bibinfo{author}{\bibfnamefont{G.~A.} \bibnamefont{Lalazissis}},
  \bibinfo{author}{\bibfnamefont{J.}~\bibnamefont{Konig}}, \bibnamefont{and}
  \bibinfo{author}{\bibfnamefont{P.}~\bibnamefont{Ring}},
  \bibinfo{journal}{Phys. Rev.} \textbf{\bibinfo{volume}{C55}},
  \bibinfo{pages}{540} (\bibinfo{year}{1997}).

\bibitem[{\citenamefont{Lalazissis et~al.}(1999)\citenamefont{Lalazissis,
  Raman, and Ring}}]{Lalazissis:1999}
\bibinfo{author}{\bibfnamefont{G.~A.} \bibnamefont{Lalazissis}},
  \bibinfo{author}{\bibfnamefont{S.}~\bibnamefont{Raman}}, \bibnamefont{and}
  \bibinfo{author}{\bibfnamefont{P.}~\bibnamefont{Ring}}, \bibinfo{journal}{At.
  Data Nucl. Data Tables} \textbf{\bibinfo{volume}{71}}, \bibinfo{pages}{1}
  (\bibinfo{year}{1999}).

\bibitem[{\citenamefont{Todd-Rutel and Piekarewicz}(2005)}]{Todd-Rutel:2005fa}
\bibinfo{author}{\bibfnamefont{B.~G.} \bibnamefont{Todd-Rutel}}
  \bibnamefont{and}
  \bibinfo{author}{\bibfnamefont{J.}~\bibnamefont{Piekarewicz}},
  \bibinfo{journal}{Phys. Rev. Lett} \textbf{\bibinfo{volume}{95}},
  \bibinfo{pages}{122501} (\bibinfo{year}{2005}).

\bibitem[{\citenamefont{Hartle}(1973)}]{Hartle:1973zza}
\bibinfo{author}{\bibfnamefont{J.~B.} \bibnamefont{Hartle}},
  \bibinfo{journal}{Astrophys. Space Sci.} \textbf{\bibinfo{volume}{24}},
  \bibinfo{pages}{385} (\bibinfo{year}{1973}).

\bibitem[{\citenamefont{Weber}(1999)}]{Weber:1999}
\bibinfo{author}{\bibfnamefont{F.}~\bibnamefont{Weber}},
  \emph{\bibinfo{title}{Pulsars as Astrophysical Laboratories for Nuclear and
  Particle Physics}} (\bibinfo{publisher}{Institute of Physics Publishing},
  \bibinfo{address}{Bristol, UK}, \bibinfo{year}{1999}).

\bibitem[{\citenamefont{Glendenning}(2000)}]{Glendenning:2000}
\bibinfo{author}{\bibfnamefont{N.~K.} \bibnamefont{Glendenning}},
  \emph{\bibinfo{title}{Compact Stars}} (\bibinfo{publisher}{Springer-Verlag
  New York}, \bibinfo{year}{2000}).

\bibitem[{\citenamefont{Landau and Lifshitz}(1975)}]{Landau:1975}
\bibinfo{author}{\bibfnamefont{L.~D.} \bibnamefont{Landau}} \bibnamefont{and}
  \bibinfo{author}{\bibfnamefont{E.~M.} \bibnamefont{Lifshitz}},
  \emph{\bibinfo{title}{The Classical Theory of Fields}}
  (\bibinfo{publisher}{Pergamon Press}, \bibinfo{address}{Oxford},
  \bibinfo{year}{1975}), \bibinfo{edition}{4th} ed.

\bibitem[{\citenamefont{Lattimer and Prakash}(2001)}]{Lattimer:2000nx}
\bibinfo{author}{\bibfnamefont{J.~M.} \bibnamefont{Lattimer}} \bibnamefont{and}
  \bibinfo{author}{\bibfnamefont{M.}~\bibnamefont{Prakash}},
  \bibinfo{journal}{Astrophys. J.} \textbf{\bibinfo{volume}{550}},
  \bibinfo{pages}{426} (\bibinfo{year}{2001}).

\bibitem[{\citenamefont{Xu et~al.}(2009{\natexlab{a}})\citenamefont{Xu, Chen,
  Li, and Ma}}]{Xu:2008vz}
\bibinfo{author}{\bibfnamefont{J.}~\bibnamefont{Xu}},
  \bibinfo{author}{\bibfnamefont{L.-W.} \bibnamefont{Chen}},
  \bibinfo{author}{\bibfnamefont{B.-A.} \bibnamefont{Li}}, \bibnamefont{and}
  \bibinfo{author}{\bibfnamefont{H.-R.} \bibnamefont{Ma}},
  \bibinfo{journal}{Phys. Rev.} \textbf{\bibinfo{volume}{C79}},
  \bibinfo{pages}{035802} (\bibinfo{year}{2009}{\natexlab{a}}).

\bibitem[{\citenamefont{Xu et~al.}(2009{\natexlab{b}})\citenamefont{Xu, Chen,
  Li, and Ma}}]{Xu:2009vi}
\bibinfo{author}{\bibfnamefont{J.}~\bibnamefont{Xu}},
  \bibinfo{author}{\bibfnamefont{L.-W.} \bibnamefont{Chen}},
  \bibinfo{author}{\bibfnamefont{B.-A.} \bibnamefont{Li}}, \bibnamefont{and}
  \bibinfo{author}{\bibfnamefont{H.-R.} \bibnamefont{Ma}},
  \bibinfo{journal}{Astrophys. J.} \textbf{\bibinfo{volume}{697}},
  \bibinfo{pages}{1549} (\bibinfo{year}{2009}{\natexlab{b}}).

\bibitem[{\citenamefont{Moustakidis et~al.}(2010)\citenamefont{Moustakidis,
  Niksic, Lalazissis, Vretenar, and Ring}}]{Moustakidis:2010zx}
\bibinfo{author}{\bibfnamefont{C.~C.} \bibnamefont{Moustakidis}},
  \bibinfo{author}{\bibfnamefont{T.}~\bibnamefont{Niksic}},
  \bibinfo{author}{\bibfnamefont{G.~A.} \bibnamefont{Lalazissis}},
  \bibinfo{author}{\bibfnamefont{D.}~\bibnamefont{Vretenar}}, \bibnamefont{and}
  \bibinfo{author}{\bibfnamefont{P.}~\bibnamefont{Ring}}
  (\bibinfo{year}{2010}), \eprint{1004.3882}.

\bibitem[{\citenamefont{Worley et~al.}(2008)\citenamefont{Worley, Krastev, and
  Li}}]{Worley:2008cb}
\bibinfo{author}{\bibfnamefont{A.}~\bibnamefont{Worley}},
  \bibinfo{author}{\bibfnamefont{P.~G.} \bibnamefont{Krastev}},
  \bibnamefont{and} \bibinfo{author}{\bibfnamefont{B.-A.} \bibnamefont{Li}}
  (\bibinfo{year}{2008}), \eprint{0801.1653}.

\bibitem[{\citenamefont{Lorenz et~al.}(1993)\citenamefont{Lorenz, Ravenhall,
  and Pethick}}]{Lorenz:1992zz}
\bibinfo{author}{\bibfnamefont{C.~P.} \bibnamefont{Lorenz}},
  \bibinfo{author}{\bibfnamefont{D.~G.} \bibnamefont{Ravenhall}},
  \bibnamefont{and} \bibinfo{author}{\bibfnamefont{C.~J.}
  \bibnamefont{Pethick}}, \bibinfo{journal}{Phys. Rev. Lett.}
  \textbf{\bibinfo{volume}{70}}, \bibinfo{pages}{379} (\bibinfo{year}{1993}).

\bibitem[{\citenamefont{{Ravenhall} and {Pethick}}(1994)}]{Ravenhall:1994}
\bibinfo{author}{\bibfnamefont{D.~G.} \bibnamefont{{Ravenhall}}}
  \bibnamefont{and} \bibinfo{author}{\bibfnamefont{C.~J.}
  \bibnamefont{{Pethick}}}, \bibinfo{journal}{\apj}
  \textbf{\bibinfo{volume}{424}}, \bibinfo{pages}{846} (\bibinfo{year}{1994}).

\bibitem[{\citenamefont{Baym et~al.}(1971)\citenamefont{Baym, Pethick, and
  Sutherland}}]{Baym:1971pw}
\bibinfo{author}{\bibfnamefont{G.}~\bibnamefont{Baym}},
  \bibinfo{author}{\bibfnamefont{C.}~\bibnamefont{Pethick}}, \bibnamefont{and}
  \bibinfo{author}{\bibfnamefont{P.}~\bibnamefont{Sutherland}},
  \bibinfo{journal}{Astrophys. J.} \textbf{\bibinfo{volume}{170}},
  \bibinfo{pages}{299} (\bibinfo{year}{1971}).

\bibitem[{\citenamefont{Roca-Maza and Piekarewicz}(2008)}]{RocaMaza:2008ja}
\bibinfo{author}{\bibfnamefont{X.}~\bibnamefont{Roca-Maza}} \bibnamefont{and}
  \bibinfo{author}{\bibfnamefont{J.}~\bibnamefont{Piekarewicz}},
  \bibinfo{journal}{Phys. Rev.} \textbf{\bibinfo{volume}{C78}},
  \bibinfo{pages}{025807} (\bibinfo{year}{2008}).

\bibitem[{\citenamefont{Ravenhall et~al.}(1983)\citenamefont{Ravenhall,
  Pethick, and Wilson}}]{Ravenhall:1983uh}
\bibinfo{author}{\bibfnamefont{D.~G.} \bibnamefont{Ravenhall}},
  \bibinfo{author}{\bibfnamefont{C.~J.} \bibnamefont{Pethick}},
  \bibnamefont{and} \bibinfo{author}{\bibfnamefont{J.~R.}
  \bibnamefont{Wilson}}, \bibinfo{journal}{Phys. Rev. Lett.}
  \textbf{\bibinfo{volume}{50}}, \bibinfo{pages}{2066} (\bibinfo{year}{1983}).

\bibitem[{\citenamefont{Hashimoto et~al.}(1984)\citenamefont{Hashimoto, Seki,
  and Yamada}}]{Hashimoto:1984}
\bibinfo{author}{\bibfnamefont{M.}~\bibnamefont{Hashimoto}},
  \bibinfo{author}{\bibfnamefont{H.}~\bibnamefont{Seki}}, \bibnamefont{and}
  \bibinfo{author}{\bibfnamefont{M.}~\bibnamefont{Yamada}},
  \bibinfo{journal}{Prog. Theor. Phys.} \textbf{\bibinfo{volume}{71}},
  \bibinfo{pages}{320} (\bibinfo{year}{1984}).

\bibitem[{\citenamefont{Horowitz
  et~al.}(2004{\natexlab{a}})\citenamefont{Horowitz, Perez-Garcia, and
  Piekarewicz}}]{Horowitz:2004yf}
\bibinfo{author}{\bibfnamefont{C.~J.} \bibnamefont{Horowitz}},
  \bibinfo{author}{\bibfnamefont{M.~A.} \bibnamefont{Perez-Garcia}},
  \bibnamefont{and}
  \bibinfo{author}{\bibfnamefont{J.}~\bibnamefont{Piekarewicz}},
  \bibinfo{journal}{Phys. Rev.} \textbf{\bibinfo{volume}{C69}},
  \bibinfo{pages}{045804} (\bibinfo{year}{2004}{\natexlab{a}}).

\bibitem[{\citenamefont{Horowitz
  et~al.}(2004{\natexlab{b}})\citenamefont{Horowitz, Perez-Garcia, Carriere,
  Berry, and Piekarewicz}}]{Horowitz:2004pv}
\bibinfo{author}{\bibfnamefont{C.~J.} \bibnamefont{Horowitz}},
  \bibinfo{author}{\bibfnamefont{M.~A.} \bibnamefont{Perez-Garcia}},
  \bibinfo{author}{\bibfnamefont{J.}~\bibnamefont{Carriere}},
  \bibinfo{author}{\bibfnamefont{D.~K.} \bibnamefont{Berry}}, \bibnamefont{and}
  \bibinfo{author}{\bibfnamefont{J.}~\bibnamefont{Piekarewicz}},
  \bibinfo{journal}{Phys. Rev.} \textbf{\bibinfo{volume}{C70}},
  \bibinfo{pages}{065806} (\bibinfo{year}{2004}{\natexlab{b}}).

\bibitem[{\citenamefont{Horowitz et~al.}(2005)\citenamefont{Horowitz,
  Perez-Garcia, Berry, and Piekarewicz}}]{Horowitz:2005zb}
\bibinfo{author}{\bibfnamefont{C.~J.} \bibnamefont{Horowitz}},
  \bibinfo{author}{\bibfnamefont{M.~A.} \bibnamefont{Perez-Garcia}},
  \bibinfo{author}{\bibfnamefont{D.~K.} \bibnamefont{Berry}}, \bibnamefont{and}
  \bibinfo{author}{\bibfnamefont{J.}~\bibnamefont{Piekarewicz}},
  \bibinfo{journal}{Phys. Rev.} \textbf{\bibinfo{volume}{C72}},
  \bibinfo{pages}{035801} (\bibinfo{year}{2005}).

\bibitem[{\citenamefont{Carriere et~al.}(2003)\citenamefont{Carriere, Horowitz,
  and Piekarewicz}}]{Carriere:2002bx}
\bibinfo{author}{\bibfnamefont{J.}~\bibnamefont{Carriere}},
  \bibinfo{author}{\bibfnamefont{C.~J.} \bibnamefont{Horowitz}},
  \bibnamefont{and}
  \bibinfo{author}{\bibfnamefont{J.}~\bibnamefont{Piekarewicz}},
  \bibinfo{journal}{Astrophys. J.} \textbf{\bibinfo{volume}{593}},
  \bibinfo{pages}{463} (\bibinfo{year}{2003}).

\bibitem[{\citenamefont{Walecka}(1974)}]{Walecka:1974qa}
\bibinfo{author}{\bibfnamefont{J.~D.} \bibnamefont{Walecka}},
  \bibinfo{journal}{Annals Phys} \textbf{\bibinfo{volume}{83}},
  \bibinfo{pages}{491} (\bibinfo{year}{1974}).

\bibitem[{\citenamefont{Serot and Walecka}(1986)}]{Serot:1984ey}
\bibinfo{author}{\bibfnamefont{B.~D.} \bibnamefont{Serot}} \bibnamefont{and}
  \bibinfo{author}{\bibfnamefont{J.~D.} \bibnamefont{Walecka}},
  \bibinfo{journal}{Adv. Nucl. Phys.} \textbf{\bibinfo{volume}{16}},
  \bibinfo{pages}{1} (\bibinfo{year}{1986}).

\bibitem[{\citenamefont{Serot and Walecka}(1997)}]{Serot:1997xg}
\bibinfo{author}{\bibfnamefont{B.~D.} \bibnamefont{Serot}} \bibnamefont{and}
  \bibinfo{author}{\bibfnamefont{J.~D.} \bibnamefont{Walecka}},
  \bibinfo{journal}{Int. J. Mod. Phys.} \textbf{\bibinfo{volume}{E6}},
  \bibinfo{pages}{515} (\bibinfo{year}{1997}).

\bibitem[{\citenamefont{Horowitz and
  Piekarewicz}(2001{\natexlab{b}})}]{Horowitz:2000xj}
\bibinfo{author}{\bibfnamefont{C.~J.} \bibnamefont{Horowitz}} \bibnamefont{and}
  \bibinfo{author}{\bibfnamefont{J.}~\bibnamefont{Piekarewicz}},
  \bibinfo{journal}{Phys. Rev. Lett.} \textbf{\bibinfo{volume}{86}},
  \bibinfo{pages}{5647} (\bibinfo{year}{2001}{\natexlab{b}}).

\bibitem[{\citenamefont{Todd and Piekarewicz}(2003)}]{Todd:2003xs}
\bibinfo{author}{\bibfnamefont{B.~G.} \bibnamefont{Todd}} \bibnamefont{and}
  \bibinfo{author}{\bibfnamefont{J.}~\bibnamefont{Piekarewicz}},
  \bibinfo{journal}{Phys. Rev.} \textbf{\bibinfo{volume}{C67}},
  \bibinfo{pages}{044317} (\bibinfo{year}{2003}).

\bibitem[{\citenamefont{Danielewicz et~al.}(2002)\citenamefont{Danielewicz,
  Lacey, and Lynch}}]{Danielewicz:2002pu}
\bibinfo{author}{\bibfnamefont{P.}~\bibnamefont{Danielewicz}},
  \bibinfo{author}{\bibfnamefont{R.}~\bibnamefont{Lacey}}, \bibnamefont{and}
  \bibinfo{author}{\bibfnamefont{W.~G.} \bibnamefont{Lynch}},
  \bibinfo{journal}{Science} \textbf{\bibinfo{volume}{298}},
  \bibinfo{pages}{1592} (\bibinfo{year}{2002}).

\bibitem[{\citenamefont{Piekarewicz}(2007)}]{Piekarewicz:2007dx}
\bibinfo{author}{\bibfnamefont{J.}~\bibnamefont{Piekarewicz}},
  \bibinfo{journal}{Phys. Rev.} \textbf{\bibinfo{volume}{C76}},
  \bibinfo{pages}{064310} (\bibinfo{year}{2007}).

\bibitem[{\citenamefont{Piekarewicz}(2004)}]{Piekarewicz:2003br}
\bibinfo{author}{\bibfnamefont{J.}~\bibnamefont{Piekarewicz}},
  \bibinfo{journal}{Phys. Rev.} \textbf{\bibinfo{volume}{C69}},
  \bibinfo{pages}{041301} (\bibinfo{year}{2004}).

\bibitem[{\citenamefont{Furnstahl}(2002)}]{Furnstahl:2001un}
\bibinfo{author}{\bibfnamefont{R.~J.} \bibnamefont{Furnstahl}},
  \bibinfo{journal}{Nucl. Phys.} \textbf{\bibinfo{volume}{A706}},
  \bibinfo{pages}{85} (\bibinfo{year}{2002}).

\bibitem[{\citenamefont{Ducoin et~al.}(2010)\citenamefont{Ducoin, Margueron,
  and Providencia}}]{Ducoin:2010as}
\bibinfo{author}{\bibfnamefont{C.}~\bibnamefont{Ducoin}},
  \bibinfo{author}{\bibfnamefont{J.}~\bibnamefont{Margueron}},
  \bibnamefont{and}
  \bibinfo{author}{\bibfnamefont{C.}~\bibnamefont{Providencia}}
  (\bibinfo{year}{2010}), \eprint{1004.5197}.

\end{thebibliography}
\end{document}